\documentclass{aa}  

\usepackage{graphicx}
\usepackage{txfonts}
\usepackage{amsmath}	% Advanced maths commands
\usepackage{amssymb}	% Extra maths symbols
\usepackage{mathrsfs}
\usepackage{float}
\usepackage{rotating}

\newcommand{\lya}{Ly$\alpha$}

\newcommand{\kms}{$\mathrm{\,km s^{-1}}$}

\newcommand{\HI}{\mbox{H\,{\sc i}}}

\newcommand{\HeII}{\mbox{He\,{\sc ii}}}
  
\begin{document}

   \title{Parameter estimation from \lya\ forest in Fourier space using Information Maximising Neural Network}

   \titlerunning{\lya\ forest - IMNN}
   \authorrunning{Maitra et al.}

   \author{Soumak Maitra
          \inst{1}\fnmsep\thanks{soumak93@gmail.com},
          Stefano Cristiani
          \inst{1},
          Matteo Viel
          \inst{2,1,3,4},
          Roberto Trotta
          \inst{2,3,4,5}
          \and
          Guido Cupani
          \inst{1}
          }

   \institute{Istituto Nazionale di Astrofisica -Osservatorio Astronomico di Trieste, Via Tiepolo 11, Trieste, Italy\\
   \and
SISSA International School for Advanced Studies, Via Bonomea 265, 34136, Trieste, Italy\\
\and
INFN - Sezione di Trieste, via Valerio 2, 34127, Trieste, Italy\\
\and
IFPU, Institute for Fundamental Physics of the Universe, Via Beirut 2, 34014 Trieste, Italy\\
\and
%$^{5}$ ICSC - Italian Research Center on High Performance Computing, Big Data and Quantum Computing, Via Magnanelli 2, Bologna, Italy\\
Imperial College London, Astrophysics Group, Physics Department, Blackett Lab, Prince Consort Road, London SW7 2AZ, UK 
             }
\date{Accepted XXX. Received YYY; in original form ZZZ}

% \abstract{}{}{}{}{} 
% 5 {} token are mandatory
 
  \abstract
   {}
   {To present a robust parameter estimation with simulated \lya\ forest spectra from Sherwood-Relics simulations suite using Information Maximizing Neural Network (IMNN) to extract maximal information from \lya\  1D-transmitted flux in Fourier space.}
   {We perform 1D estimations using IMNN for Intergalactic Medium (IGM) thermal parameters $T_0$ and $\gamma$, at $z=2-4$, and cosmological parameters $\sigma_8$ and $n_{\rm s}$ at $z=3-4$. We compare our results with estimates from power spectrum using the posterior distribution from a Markov Chain Monte Carlo (MCMC). We then check the robustness of IMNN estimates against deviation in spectral noise levels, continuum uncertainties, and instrumental smoothing effects. Using mock \lya\ forest sightlines from publicly available CAMELS project, we also check the robustness of the trained IMNN on a different simulation.  We also perform a 2D-parameter estimation for $T_0$ and \HI\ photo-ionization rates $\Gamma_{\rm HI}$.}
   {We obtain improved estimates of $T_0$ and $\gamma$ using IMNN over standard MCMC approach. These estimates are also more robust against SNR deviations at $z=2$ and 3. At $z=4$, the sensitivity to noise deviations is on par with MCMC estimates. The IMNN also provides $T_0$ and $\gamma$ estimates which are robust against continuum uncertainties by extracting continuum-independent small-scale information from the Fourier domain. In case of $\sigma_8$ and $n_{\rm s}$, the IMNN performs on par with MCMC but still offers a significant speed boost in estimating parameters from a new dataset. 
The improved estimates with IMNN are seen for high instrumental-resolution (FWHM=6\kms). At medium or low resolutions the IMNN performs similar to MCMC, suggesting an improved extraction of small-scale information with IMNN. We also find that IMNN estimates are robust against the choice of simulation. By performing a 2D-parameter estimation for $T_0$ and $\Gamma_{\rm HI}$ we also demonstrate how to take forward this approach observationally in the future.}
   {}


   \maketitle
%
%-------------------------------------------------------------------

\section{Introduction}
The \lya\ forest absorption spectra consist of numerous absorption lines created by the $1s\rightarrow 2p$ \lya\ transition of the intervening neutral hydrogen gas along the lines of sight to distant quasars. These absorption features trace the underlying distribution of matter in the universe, thus revealing the cosmic web's intricate filamentary structure \citep{finley2014,lee2018}. The large-scale structure traced by the \lya\ forest is sensitive to the nature of dark matter and dark energy and allows us to test structure formation processes down to small scales. By studying the clustering and distribution of matter on cosmic scales, one can place constraints on the density fluctuations of dark matter \citep{croft1998, croft1999, mcdonald2000, croft2002, mcdonald2003,viel2004b} and the equation of state of dark energy \citep{viel2003,coughlin2019}, along with cosmological parameters \citep{viel2004a,mcdonald2000,viel2006b}, warm dark matter models \citep{viel2013,irsic2017}, neutrino mass \citep{palanque2015a,palanque2015b,yeche2017,palanque2020}, etc. governing the constitution of the Universe. In the astrophysical context, \lya\ forest is also useful in constraining the Intergalactic Medium (IGM) temperature $T_0$ at cosmic mean density and slope $\gamma$ of the temperature ($T$)–density ($\Delta$) relation ($T=T_0(\Delta)^{\gamma-1}$)\citep{schaye1999,schaye2000,theuns2000b,mcdonald2001,becker2011,boera2014,gaikwad2021}, cosmic reionization \citep{fan2006,worseck2018} and the impact of various feedbacks processes (such as SNe and AGN driven outflows) on the IGM that operates during the formation and evolution of galaxies over cosmic time \citep{aguirre2001,oppenheimer2006}.

The \lya\ forest can also act as a complementary source of information to other cosmological probes like the cosmic microwave background (CMB) and galaxy surveys. Combining data from various sources, including the Lyman forest, allows for more robust and precise cosmological parameter estimation, reducing potential biases and uncertainties \citep{viel2004c,viel2006b,lesgourgues2007s}.
In order to extract critical information about the large-scale structure, BAO, redshift space distortions, and the nature of cosmic density fluctuations, astrophysicists have investigated the clustering of \lya\ forest, customarily adopting power Spectrum of the transmitted flux as a summary statistics \citep{mcdonald2000,mcdonald2006,croft2002,seljak2006}. The advent of high-fidelity spectra also allows one to perform higher-order clustering studies with \lya\ forest \citep{viel2004a,tie2019,maitra2019,maitra2022a,maitra2022b}.

Moving forward,
machine learning approaches and neural networks (NN) can enhance the precision and accuracy of parameter estimates from relevant observables, potentially leading to more reliable astrophysical and cosmological models \citep{gupta2018,charnock2018,ribli2019,nayak2023}.  Additionally, a trained neural network also offers a substantial boost in computational efficiency when dealing with new datasets \citep{nygaard2023}. This feature becomes especially important when estimating parameters from the current age of large astrophysical and cosmological datasets. Parameter estimation using neural networks on the real space field information can offer significant constraints on the estimated parameters. However, since neural networks are exceptional at picking up non-linear features in the training data, this makes them very susceptible to simulation-specific features in the training data \citep[see][for example]{gluck2023}. Thus, robustness against learning unwanted features in the training data can become an issue for inference approaches done from the real space field. 

An alternative approach can be to work in the Fourier domain to filter out such simulation-specific non-linearities in the real space and draw only relevant information out from the Fourier space using neural networks. In the past, neural network approaches have been used to estimate parameters from the Fourier space that were more accurate and robust in comparison to traditional maximum likelihood approaches from the power spectrum \citep[see][for example]{navarro2022}. Some of these applications have also focussed on the \lya\ forest to extract thermal parameters \citep{alsing18} or detect large-scale power enhancement in the power spectrum due to patchy reionization \citep{molaro2022,molaro23}. Here, we work on extracting maximal information out of the data in the Fourier space using neural networks and then compare this approach with traditional maximum likelihood approaches using the power spectrum. In particular, we use Information Maximizing Neural Networks \citep[IMNN,][]{charnock2018} to extract maximal information from the Fourier space data and enhance the parameter estimation.  Working in Fourier space, while reducing the data dimensionality also retains relevant information regarding the clustering of \lya\ forest. We then compare our results with a traditional parameter estimation using a maximum likelihood approach from the flux power spectrum. 

In this work, we mainly focus on 1D parameter estimation and its robustness for the thermal parameters $T_0$ and $\gamma$, and cosmological parameters $\sigma_8$ and $n_S$, individually. We find that training the neural networks for N-D parameter estimation requires simulations having simultaneous parameter variations \citep[similar to][]{nayak2023} which we do not have access to right now. However, we perform a 2D parameter estimation for $T_0$ and \HI\ photo-ionization rate $\Gamma_{\rm HI}$ since $\Gamma_{\rm HI}$ can be varied simply as a post-processing step during the forward modeling of mock \lya\ forest spectra without the need to run a new simulation. This is done to demonstrate how to take forward the IMNN approach in the future observationally for a full N-D parameter estimation. While such endeavors will require a vast array of simulations for training the neural network, one can draw inspiration from recent works related to generating significantly efficient simulated training examples.  For example, one way to provide the necessary training examples would be to use fast emulators, like the recently released 21cmEMU~\citep{Breitman:2023pcj} which is able to produce summary statistics from a full run of 21cmFAST with a speed-up factor of $\sim 10^4$. Other promising approaches to producing cheap simulations rely on Lagrangian Deep Learning \citep{Dai:2020ekz} and its successors \citep[][to appear]{Rigo2024}.

\section{Simulations and mock sightlines}

For this work, we use the Sherwood-Relics suite of hydrodynamical simulations with box size (40 $h^{-1}$ cMpc)$^3$ and having $2\times 1024^3$ particles \citep[][]{puchwein2023,bolton17} to generate mock \lya\ forest sightlines to train the neural network for parameter estimation. 
The fiducial simulation used for this work was run with a standard $\Lambda$CDM cosmology with cosmological parameters based on \cite{planck2014} (\{$\Omega_m$, $\Omega_b$, $\Omega_\Lambda$, $\sigma_8$, $n_{\rm s} $, $h$\} = \{0.308, 0.0482, 0.692, 0.829, 0.961, 0.678\}). We then use simulations with varied cosmological and astrophysical parameters to train the neural network to learn from the associated variations in \lya\ forest and perform parameter estimation. We perform parameter estimation for the cosmological parameters $\sigma_8$ and $n_{\rm s} $, and the astrophysical parameters $T_0$ and $\gamma$ individually. $T_0$ is the IGM temperature at mean cosmic density and $\gamma$ is the slope of the IGM temperature $T$ and density $\Delta$ relation ($T=T_0 \Delta^{\gamma-1}$). The fiducial simulation has $\sigma_8$ and $n_{\rm s} $ values of 0.829 and 0.961, as mentioned earlier, and a redshift independent $\gamma$ of 1.3. The temperature evolves with redshift. In the case of $\sigma_8$, simulations were used with variation in the parameter as $\sigma_8=0.829\pm 0.075$. For $n_{\rm s} $, parameter variations used were $n_{\rm s} =0.961\pm 0.04$. In the case of $T_0$, we use simulations having temperatures 1.5 times higher and lower than the fiducial simulation. For $\gamma$, simulations having $\gamma=1.3\pm 0.3$ were used. In total, we use 9 simulations to train the neural network to learn parameter estimation for all these parameters individually. It might be worth mentioning here that all these simulations with variations in parameters were run with the same initial seed density fields. We also use 4 other simulations having $\sigma_8=0.804, 0.854$ and $n_{\rm s} =0.941, 0.981$. These simulations were used to test the ability of the neural network to predict parameter values different from what it has been trained on. Additionally, we also use 3 other simulations run with fiducial parameters, but with different initial seed density fields on which we perform the testing of the parameter estimation. This is done to check how robust is the trained neural network against new unseen data. All the above simulations mentioned were run with spatially homogenous photo-ionization rates and photo-heating rates from the fiducial UV background model presented in \cite{puchwein2019} (see Table D1). For the 2D parameter estimation of $T_0$ and \HI\ photo-ionization rate $\Gamma_{\rm HI}$, we vary $\Gamma_{\rm HI}$ by 25\% as a post-processing step on \lya\ forest spectra by uniformly scaling the optical depth field $\tau$. The list of all the simulations used in this work is tabulated in Table.~\ref{Tab:Sims}.

 \begin{table*}
 \centering

\caption{List of Sherwood-Relics simulations used }
\begin{tabular}{llllllll}
\hline
\textbf{Simulation Model} & \textbf{Purpose} & \textbf{Seed} & \textbf{Temperature}  & \textbf{$\gamma$}  & \textbf{$\sigma_8$} & \textbf{$n_{\rm s} $} & \textbf{$\Gamma_{\rm HI}$}\\
\hline
\\
40-1024 (Fiducial) & Training & 181170 & Fiducial  & 1.3 &  0.829 & 0.961 & \cite{puchwein2019} \\
\\
40-1024-cold & Training & 181170 & Fiducial/1.5  & 1.3 &  0.829 &
0.961 & \cite{puchwein2019} \\
40-1024-hot & Training & 181170 & Fiducial*1.5  & 1.3 &  0.829 & 0.961
 & \cite{puchwein2019} \\
\\
40-1024-g10 & Training & 181170 & Fiducial  & 1.0 &  0.829 & 0.961
 & \cite{puchwein2019} \\
40-1024-g16 & Training & 181170 & Fiducial  & 1.6 &  0.829 & 0.961
 & \cite{puchwein2019} \\
\\
40-1024-s754 & Training & 181170 & Fiducial  & 1.3 &  0.754 & 0.961
 & \cite{puchwein2019} \\
40-1024-s804 & Training & 181170 & Fiducial  & 1.3 &  0.804 & 0.961
 & \cite{puchwein2019} \\
40-1024-s854 & Training & 181170 & Fiducial  & 1.3 &  0.854 & 0.961
 & \cite{puchwein2019} \\
40-1024-s904 & Training & 181170 & Fiducial  & 1.3 &  0.904 & 0.961
 & \cite{puchwein2019} \\
\\
40-1024-n921 & Training & 181170 & Fiducial  & 1.3 &  0.829 & 0.921
 & \cite{puchwein2019} \\
40-1024-n941 & Training & 181170 & Fiducial  & 1.3 &  0.829 & 0.941
 & \cite{puchwein2019} \\
40-1024-n981 & Training & 181170 & Fiducial  & 1.3 &  0.829 & 0.981
 & \cite{puchwein2019} \\
40-1024-n1001 & Training & 181170 & Fiducial  & 1.3 &  0.829 & 1.001
 & \cite{puchwein2019} \\
 \\
40-1024 (lowered $\Gamma_{\rm HI}$) & Training & 181170 & Fiducial  & 1.3 &  0.829 & 0.961 & \cite{puchwein2019}/1.25 \\
40-1024 (elevated $\Gamma_{\rm HI}$) & Training & 181170 & Fiducial  & 1.3 &  0.829 & 0.961 & \cite{puchwein2019}*1.25 \\
\\
40-1024-seed001 & Testing & 965431 & Fiducial  & 1.3 &  0.829 &  0.961
 & \cite{puchwein2019} \\
40-1024-seed002 & Testing & 126642 & Fiducial  & 1.3 &  0.829 & 0.961
 & \cite{puchwein2019} \\
40-1024-seed003 & Testing & 140516 & Fiducial  & 1.3 &  0.829 & 0.961
 & \cite{puchwein2019} \\
\\
\hline
 
\end{tabular}
\label{Tab:Sims}
\end{table*}

The snapshots for all these simulations are stored at redshift intervals of $\Delta z=0.1$. 
We create mock \lya\ forest transmitted flux sightlines from these simulations of length 40 $h^{-1}$ cMpc at redshifts $z=2.0, 3.0$, and 4.0. This is done to test parameter estimation and the constraining power of \lya\ forest at different redshifts.
We generate the sightlines by first gridding 40$h^{-1}$cMpc (length of each sightline) into 2048 grids in wavelength. For the sake of simplicity and the fact that we are not comparing the simulated spectra with observations in this work, we retain the uniform gridding in $\Delta x=40/2048 h^{-1}$cMpc (corresponding to $\Delta z\approx 1.96\times 10^{-5}$) inherent in the simulations. We have also adjusted the photo-ionizing background for each of the varied simulations so that their mean flux matches the fiducial one. To check the effect of instrumental smoothing on the estimation, we have also generated sightlines convolved with Gaussian profiles with FWHM=6, 50, and 150 \kms. However, unless otherwise mentioned, we used transmitted flux without any Gaussian convolution. Additionally, we also add random Gaussian noise to the transmitted flux. We use uniform SNR along a single sightline but vary the SNR levels between different sightlines. The SNR values corresponding to each sightline is drawn uniformly in the logarithmic space in SNR, in the range $\log \text{SNR}\in [\log 20, \log 100]$.  This ensures that our sample has a larger number of low-SNR sightlines, similar to the observed spectra. We refer to this SNR distribution as SNR$_{Fid}$ (for ``fiducial'') from now on. We also generate sightlines having $0.85\times$SNR$_{Fid}$ to evaluate the sensitivity of the parameter estimation to deviations (in this case, systematic suppression of SNR) in noise levels. 

\iffalse
The effects of the variations of $T_0, \gamma, \sigma_8$ on the simulated \lya\ forest mock spectra at $z=2,3,4$ have been shown in Fig.~\ref{Fig:Flux}. In the case of $T_0$ and $\gamma$, the effects are primarily related to the thermal broadening of absorption features. For $T_0$, the broadening behavior is simpler since the temperature field is being scaled uniformly across the sightline. Varying $\gamma$ introduces a density-dependent scaling of the temperature field that results in thermal broadening effects that depend on the underlying density fields. In the case of $\sigma_8$, the behavior is more complicated. Since it varies the fluctuations in the underlying density fields, we not only see differences in the strength of absorptions but also redshift space shifting of these absorption features. These effects are stronger in the case of stronger absorptions suggesting that the peculiar velocity effects, which cause the stronger absorption features (tracing larger density fields) to shift more also play an important role when one varies $\sigma_8$. Thus, it would be interesting to see how a neural network-based parameter estimation performs relative to more traditional power spectrum based parameter estimation in estimating $T_0, \gamma, \sigma_8$ that affects the \lya\ forest at various levels of complications.
\fi
 
\section{Parameter estimation with MCMC from power spectrum}\label{Sec:MCMC}

For the traditional approach, we use the 1D flux power spectrum $P_F(k)$ computed over mock \lya\ forest spectra of length 40$h^{-1}$cMpc to estimate $T_0, \gamma$ and $\sigma_8$ using maximum likelihood approach. To compute the 1D flux power spectrum, we first Fourier tranform the 1D flux deviation field $\delta_F(x)=(F(x)-\bar{F})/\bar{F}$ to $\delta_F(k)$. The power spectrum is then simply proportional to $|\delta_F(k)|^2$. The 1D flux power spectrum is then normalized as
\begin{equation}
    \sigma_F^2=\int^{\infty}_{-\infty}\frac{dk\, P_F(k)}{2\pi},
\end{equation}
where $\sigma_F^2$ is the variance of the field $\delta_F(k)$. We then bin the power spectra in 10 equispaced logarithmic bins in $k$ ranging from 0.314 to 31.4 $h$cKpc$^{-1}$. The smallest scales here correspond to 100$h^{-1}$ckpc and the largest scale corresponds to 10$h^{-1}$cMpc. We intentionally went to very small scales to allow the neural network described in the following section to extract maximal information from such scales which are known to be sensitive to noise.
\begin{figure*}
\centering
    \includegraphics[viewport=0 40 300 260,width=8.5cm, clip=true]{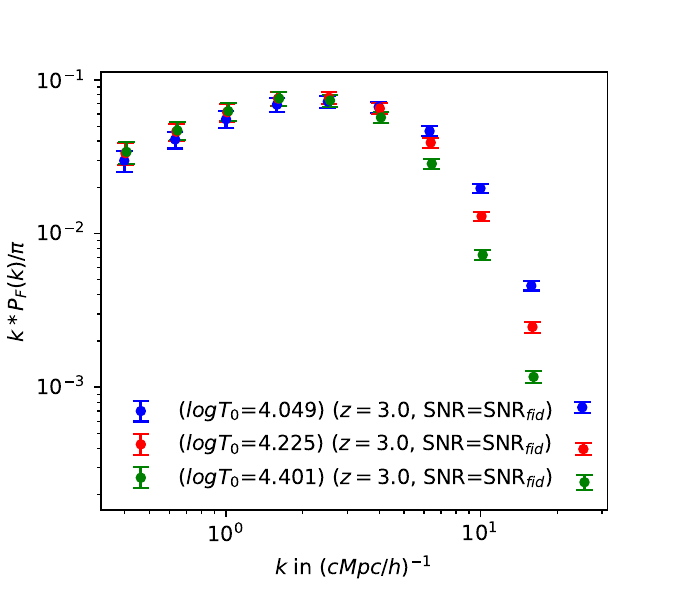}%
    \includegraphics[viewport=40 40 300 260,width=7.5cm, clip=true]{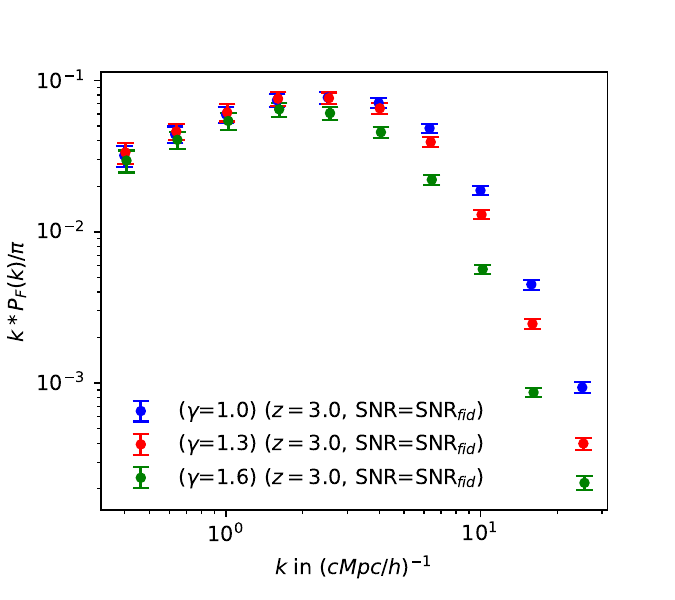}%
    
    \includegraphics[viewport=0 40 300 260,width=8.5cm, clip=true]{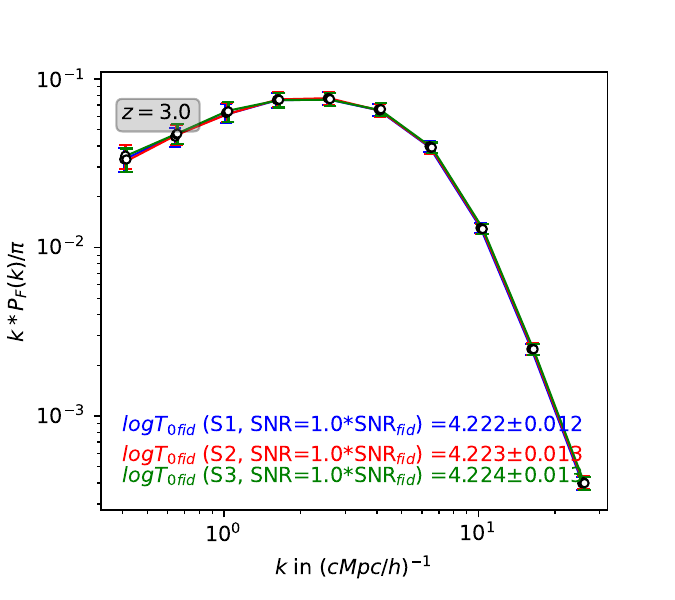}%
    \includegraphics[viewport=40 40 300 260,width=7.5cm, clip=true]{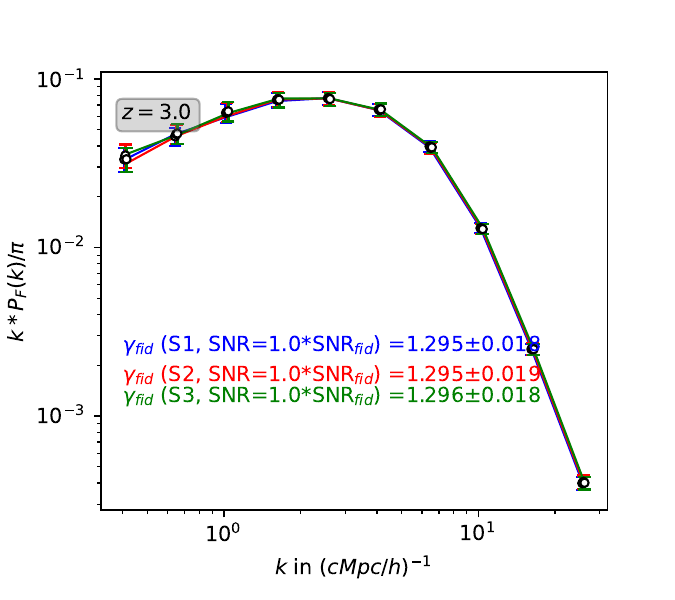}%

    \includegraphics[viewport=0 0 300 260,width=8.5cm, clip=true]{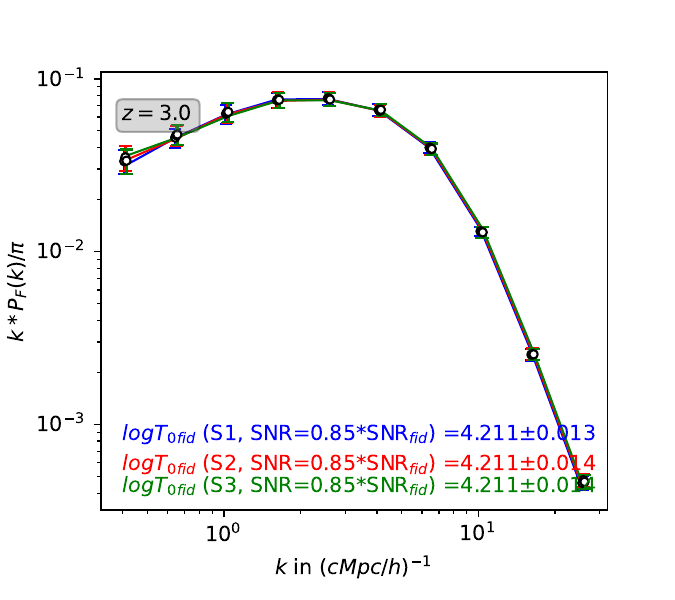}%
    \includegraphics[viewport=40 0 300 260,width=7.5cm, clip=true]{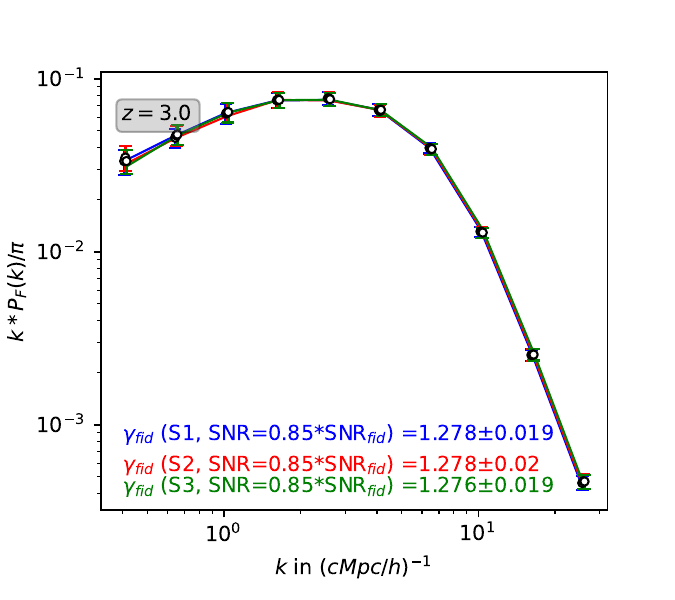}%
    
\caption{The top panels show the plots of \lya\ forest flux power spectrum corresponding to fiducial and varied astrophysical parameters $T_0$ and $\gamma$ from the \lya\ forest flux power spectrum. We have added Gaussian noise to the sightlines with an SNR distribution ranging from 20 to 100 (SNR distribution is uniform in the log scale making the distribution have more low SNR sightlines). We plot the power spectrum corresponding to the fiducial simulation (red) and the varied parameters (blue and green) corresponding to mock \lya\ with this SNR distribution. The errorbars correspond to bootstrapping errors computed over 5000 sightlines for a sample size of 50 sightlines. The power spectrum corresponding to the simulation runs (with similar SNR distribution) with fiducial parameters but different initial seed density fields are plotted with colored hollow points in the middle panels.  We model the power spectrum based on the curves in the top panels. We then use the posterior distribution by running a Markov Chain Monte Carlo (MCMC) with flat priors (see Eq.~\ref{Eq:Prior}) to estimate the parameter values (see Sec.~\ref{Sec:MCMC}). In the middle panels, blue, green, and red hollow points correspond to the fiducial simulations with different seeds, and the posterior estimate of the parameters (and its posterior standard deviation) are given in the plots. In the bottom panels, we show the sensitivity of the parameter estimates based on power spectrum to noise levels when we lower the SNR distribution ($0.85\times \rm SNR_{fid}$) of the sightlines and use the same power spectrum modeling to estimate the parameters.}
\label{Fig:PS}
\end{figure*}

\begin{figure*}
\centering
    \includegraphics[viewport=0 37 300 260,width=8.5cm, clip=true]{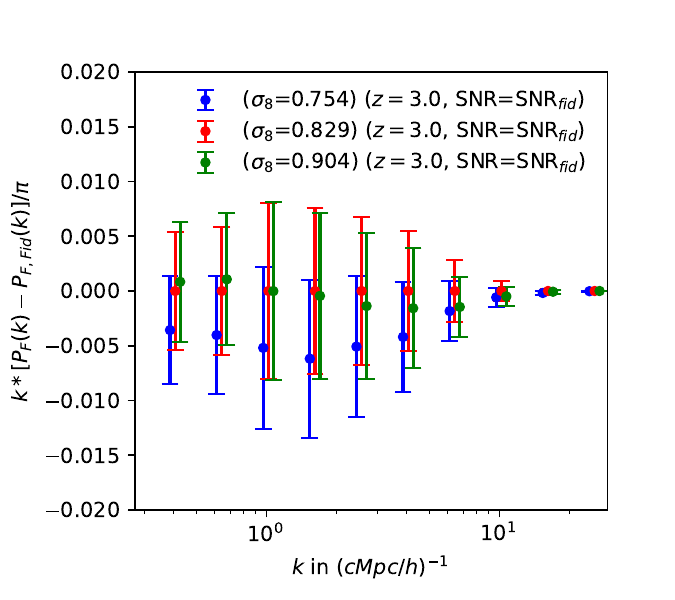}%
    \includegraphics[viewport=37 37 300 260,width=7.5cm, clip=true]{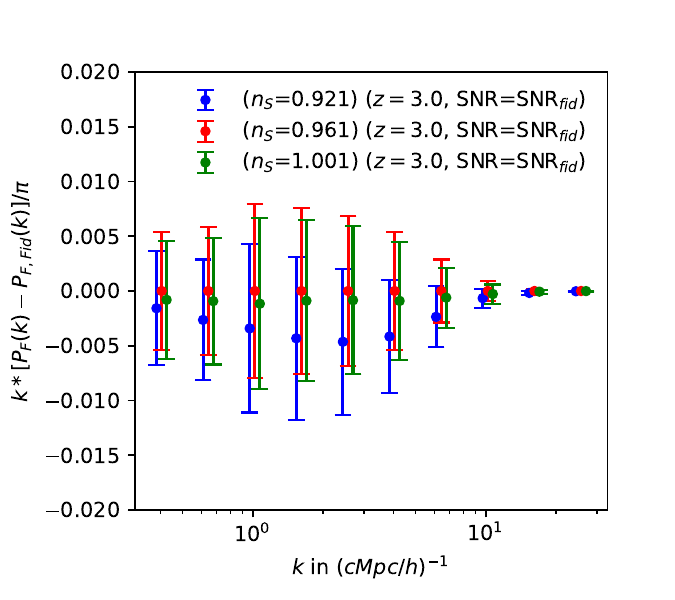}%
    
    \includegraphics[viewport=0 40 300 260,width=8.5cm, clip=true]{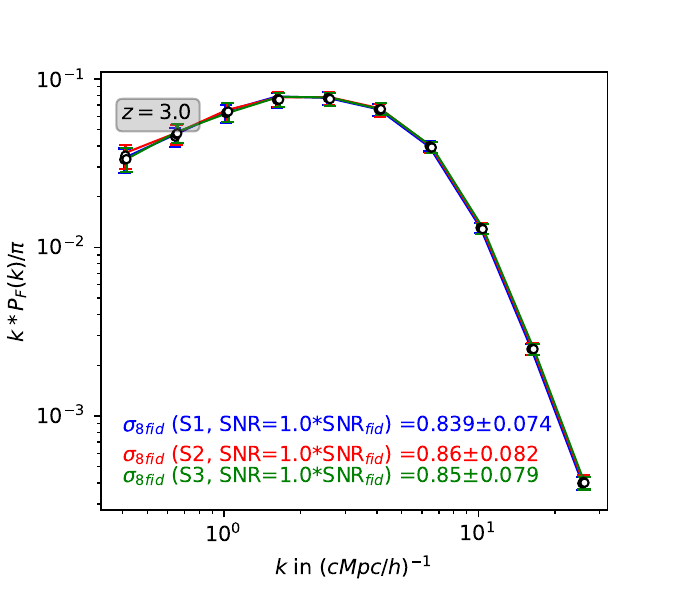}%
    \includegraphics[viewport=40 40 300 260,width=7.5cm, clip=true]{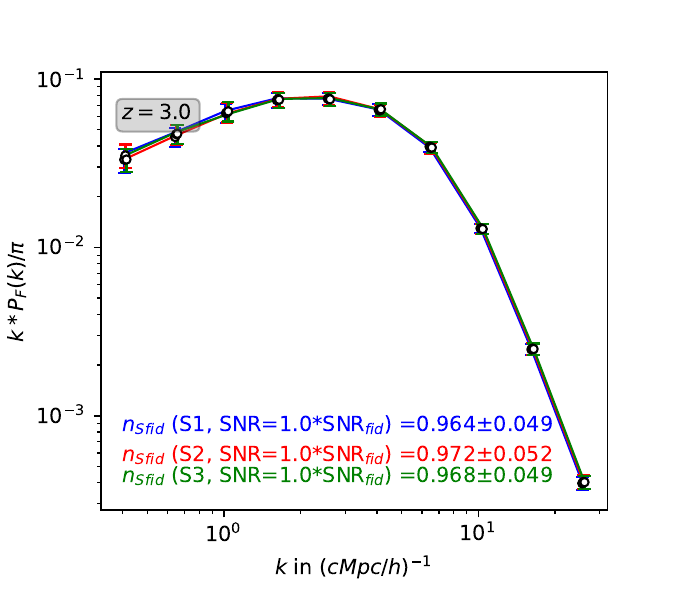}%

    \includegraphics[viewport=0 0 300 260,width=8.5cm, clip=true]{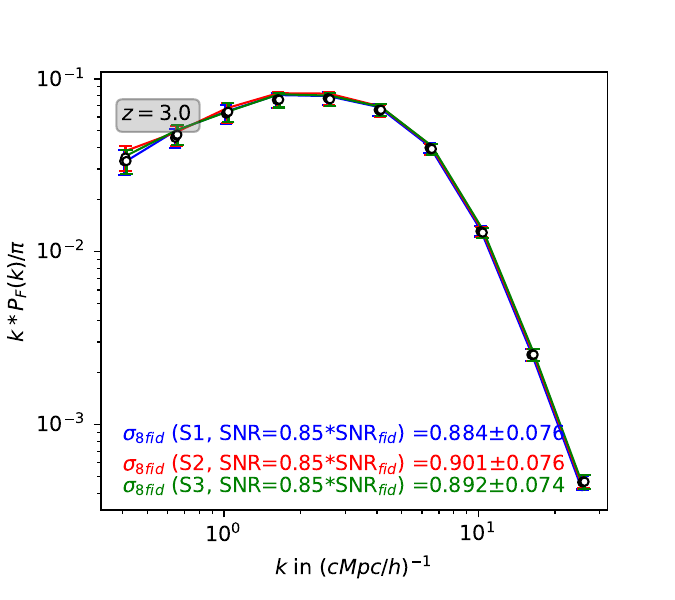}%
    \includegraphics[viewport=40 0 300 260,width=7.5cm, clip=true]{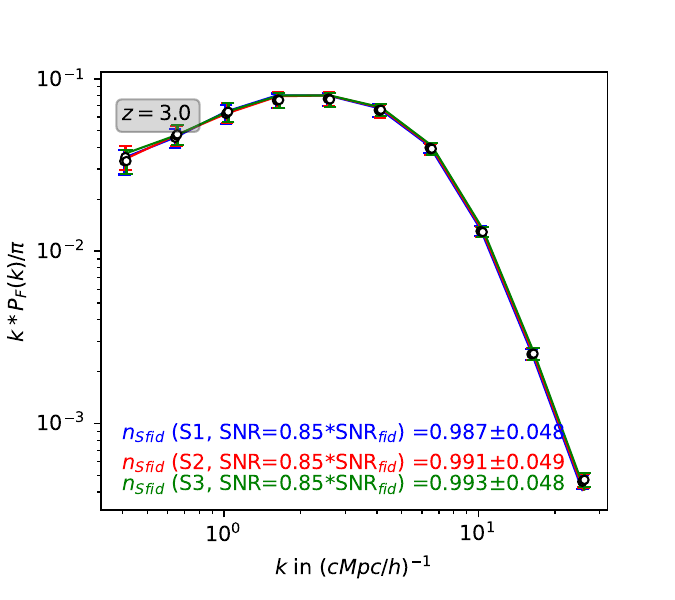}%
    
\caption{Same as Fig.~\ref{Fig:PS} but with cosmological parameters $\sigma_8$ and $n_{\rm s} $. In the top panels, however, we plot the difference between the power spectrum at a certain parameter value to the one at the fiducial parameter. This is done for a better visualization since the variations in $\sigma_8$ and $n_S$ do not cause an appreciable change in the power spectrum within the errorbars shown. }
\label{Fig:PS2}
\end{figure*}

We compute the $P_F(k)$s corresponding to the simulation at fiducial parameter values and also, for the simulations with parameter variation. The $P_F(k)$s corresponding to the fiducial and varied astrophysical parameters ($T_0$ and $\gamma$) are shown in the top panels of Fig.~\ref{Fig:PS}. The errorbars shown in the figure are bootstrap errorbars computed over 5000 sightlines for a sample size of 50 sightlines. In Fig.~\ref{Fig:PS2}, we plot the difference of the $P_F(k)$ with respect to the fiducial $P_F(k)$ for variations in the cosmological parameters ($\sigma_8$ and $n_S$). This is done to highlight the differences better since the variations in the cosmological parameters do not cause appreciable changes in the power spectrum with respect to the errorbars shown. Now, to obtain a model for $P_F(k)$ and its variation with the parameters, we approximate the power spectrum by a Taylor expansion to first order around the fiducial parameter value $\vartheta_0$ and then model $P_F(k)$ for the varied parameter $\vartheta$ about the fiducial parameter as \citep[check Sec.~4.1 of][for description]{viel2006a}
\begin{equation}\label{Eq:PS_model}
P_F(k,\vartheta)=P_F(k,\vartheta)+\frac{\partial P_F(k,\vartheta)}{\partial \vartheta}\Big|_{\vartheta=\vartheta_0}  (\vartheta-\vartheta_0).
\end{equation}
Using this, we develop a model to capture the linear variation of $P_F(k)$ with parameters $T_0, \gamma$, $\sigma_8$, and $n_S$ individually. This modeling has been done corresponding to sightlines having noise levels drawn from SNR$_{Fid}$. We then define a log-likelihood function of the form:
\begin{equation}
-2 \log \mathcal{L}(\vartheta)= ({\bf P}_F(\bf{k}, \vartheta)- \bf{d}_0)^T {C^{-1}}({\bf P}_F(\bf{k}, \vartheta)- \bf{d}_0)
\end{equation}
where $\bf{d}_0 = {\bf P}_F(\bf{k}, \vartheta_0)$ is the power spectrum values for the fiducial parameters,  arranged in a 10-dimensional vector binned over $k$-values in the range 0.314 to 31.4 $h$cKpc$^{-1}$   
 and $\bf C$ is the covariance matrix obtained from the bootstrap realizations of the power spectrum with a sample size of 50 sightlines, as mentioned before. Using this log-likelihood function, we then use Markov Chain Monte Carlo (MCMC) with uniform priors of the form
\begin{equation}\label{Eq:Prior}
    p(\vartheta) = 
    \begin{cases}
        (5 \times \Delta \vartheta)^{-1} & \text{if } |\vartheta - \vartheta_0| \leq 5\times\Delta \vartheta \\
        0 & \text{if } |\vartheta - \vartheta_0| > 5\times \Delta \vartheta
    \end{cases}
\end{equation}
(where $\Delta \vartheta$ is the variation in the model parameters for the simulations with varied parameters) to estimate the parameter values from a new set of 3 simulations run with different initial random seeds at the fiducial parameter values. We use the python package {\sc emcee} \citep{emcee2013} for running MCMC. In the middle panels of Fig.~\ref{Fig:PS} and Fig.~\ref{Fig:PS2}, we show the estimated parameter values from these 3 simulations. This has been done using sightlines whose noise levels have been drawn from SNR$_{Fid}$. To check the sensitivity of the parameter estimation to the noise levels, we then use the $P_F(k)$ modeling done using SNR$_{Fid}$ sightlines to estimate parameters for $0.85\times$SNR$_{Fid}$ sightlines. We show the estimated parameters in the bottom panels in Fig.~\ref{Fig:PS}. We subsequently compare these results with the neural network approach in Sec.~\ref{Sec:Param_estimate}. It is to be noted that we use Eq.~\ref{Eq:PS_model} for modeling individual parameter variations for 1D parameter estimations. In Sec.~\ref{Sec:Param_2d}, where we demonstrate a 2D parameter estimation for $T_0$ and $\Gamma_{\rm HI}$ we model the power spectrum for parameter variations by following a linear interpolation scheme between different simulations.

\section{Information Maximizing Neural Network}

Summarizing large datasets into a collection of sufficient summary statistics (mean flux, power spectrum, etc. for example) is becoming a necessary approach to deal with current cosmological and astronomical data. The aim is to reduce the data into the least number of summary statistics with minimum loss of information. The Massively Optimised Parameter Estimation and Data (MOPED; \cite{heavens2000}) is a popular approach of summarizing the data, wherein the summaries are the linear combinations of the data reducing the number of data points down to the number of model parameters describing the data. MOPED is entirely lossless under the assumption that the noise is independent of the model parameter and the likelihood, at least up to the first approximation, is Gaussian. However, using a linear combination of the data for the compression might not be the most optimal approach. Using machine learning, the Information Maximizing Neural Network (IMNN) provides a more convenient and informative way to compress the data into non-linear summaries \citep[check][for a recent work showing the constraining power of several 21cm summary statistics using IMNN]{prelogovic2024}.

Drawing motivation from the MOPED algorithm, IMNN aims to find some transformation $f: {\bf d} \rightarrow {\bf x}$ which maps the data ($\bf d$) to the compressed summary ($x_{\alpha}$, for the model parameter $\alpha$) \citep[check][for reference]{charnock2018}. 

It transforms the original likelihood into the form 
\begin{equation}
    -2 ln \mathscr{L} ({\bf x| \vartheta}) = { ({\bf x}-\mu_f(\vartheta))^T {\bf C}_f^{-1} ({\bf x}-\mu_f(\vartheta))} 
\end{equation}
where
\begin{equation}
    \mu_f(\vartheta)=\frac{1}{n_s}\sum_{i=1}^{n_s}{\bf x}_i^s
\end{equation}
is the mean of $n_s$ summaries. $\vartheta$ is the set of model parameters and $C_f$ is the covariance matrix. The modified Fisher information matrix can be expressed as 
\begin{equation}\label{Eq:Fisher}
    {\bf F}_{\alpha\beta}= Tr[\mu_{f,\alpha}^T {\bf C}_f^{-1}\mu_{f,\beta}]
\end{equation}
where $\mu_{f,\alpha}$ is the partial derivative of $\mu_f$ with respect to the model parameter. Since the model parameters appear only in simulations, numerical differentiation is done to compute this. The numerical differentiation is performed using three different simulations, one at the fiducial parameter value and the other two at some small deviations from the fiducial parameter. We then use a neural network to find this mapping function with the Fisher information as the reward function (maximizes the Fisher information). This ensures a mapping that preserves maximal information.
After this mapping, one can then get model parameter estimates $\vartheta_\alpha$ for the compressed testing data using score-compression,
\begin{equation}   \label{Eq:Estimate} {\vartheta}_\alpha=\vartheta^{\rm{fid}}_\alpha+{\bf F}^{-1}_{\alpha\beta}\mu_{f,\beta}^T{\bf C}^{-1}(x-\mu_f) .
\end{equation}

In this work, we use the IMNN approach to extract maximal information out of \lya\ forest transmitted flux in the Fourier domain and perform 1D model parameter estimation on astrophysical parameters $T_0$ and $\gamma$, and cosmological parameter $\sigma_8$ and $n_S$. We do this individually for each parameter using 3 simulations for each of them since training the IMNN for an N-D parameter estimation requires training sets where the parameters are varied simultaneously and we currently do not have access to such simulations (In Sec.~\ref{Sec:Param_2d}, we demonstrate a 2D parameter estimation case for $T_0$ and $\Gamma_{\rm HI}$, where $\Gamma_{\rm HI}$ has been varied for all $T_0$ values as a post-processing step without the need to run new simulations). Using IMNN, we first compress the training set to a summary statistic and then use it to get parameter estimates from the testing data set using Eq.~\ref{Eq:Estimate}.

\begin{figure*}
\centering
$T_0$ \hspace{7.5cm} $\gamma$ 
    \includegraphics[viewport=35 0 305 200,width=7.5cm, clip=true]{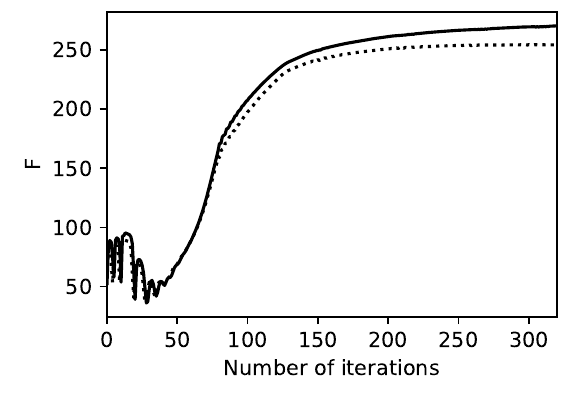}%    
    \includegraphics[viewport=35 0 305 200,width=7.5cm, clip=true]{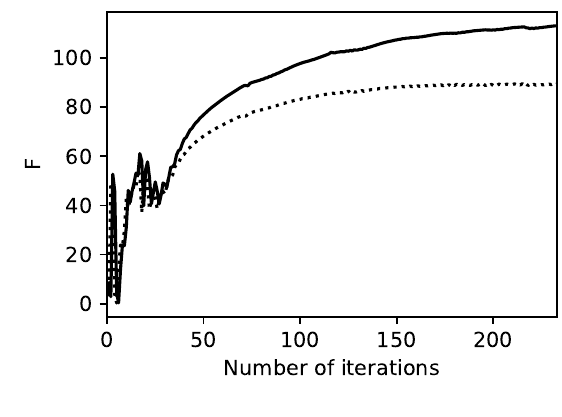}%

$\sigma_8$ \hspace{7.5cm} $n_{\rm s} $ 
\includegraphics[viewport=35 0 305 200,width=7.5cm, clip=true]{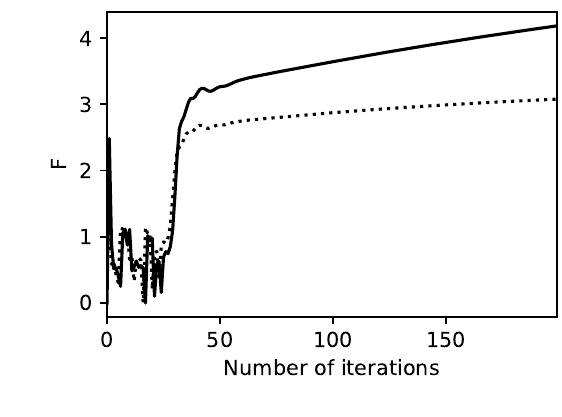}%
\includegraphics[viewport=35 0 305 200,width=7.5cm, clip=true]{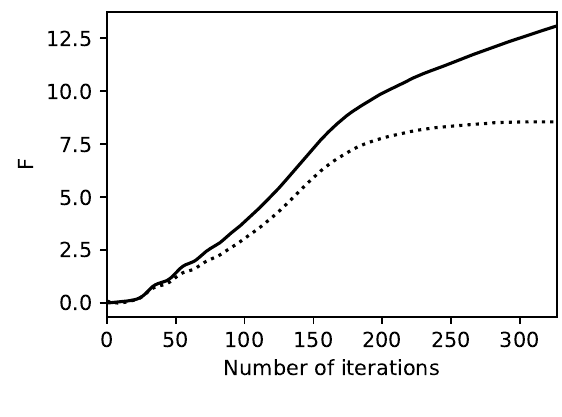}%
\caption{IMNN training example (using NumericalGradientIMNN subclass of IMNN) corresponding to the astrophysical parameters $T_0$ and $\gamma$ (top panels), and cosmological parameters $\sigma_8$ and $n_{\rm s} $ (bottom panels) for $z=3.0$.  The plots show the evolution of Fisher information during training with the training epoch. The solid lines correspond to the training sample and the dotted lines correspond to the validation sample. }
\label{Fig:Training}
\end{figure*}

\section{Training the neural network}
    
   For 1D parameter estimation of $T_0$, $\gamma$, $\sigma_8$ and $n_S$, we will use the NumericalGradientIMNN\footnote{\url{https://docs.aquila-consortium.org/imnn/main/pages/examples/subclasses/NumericalGradientIMNN/NumericalGradientIMNN.html}} subclass of IMNN which uses the derivatives of the network outputs with respect to the physical model parameters necessary to fit an IMNN. The simulations at parameter values above and below the fiducial values are used for this. To train the IMNN, we generate 5000 sightlines in total. We use 3000 of these as the training set for the neural network. The rest 2000 sightlines are used as the validation set for the training. The validation set is used at each training epoch to validate the neural network trained on the training set and then adjust the network hyperparameters accordingly. As an input to the IMNN, we use the 1D field $\sqrt{k}\delta_F(k)$, where $\delta_F(k)$ is the Fourier transformed flux deviation field and $k$ is the corresponding wave number, ranging linearly from 0.314 to 31.4$h$kpc (the same range as for the power spectrum analysis using MCMC) and containing 197 entries. 

We use 3 different simulations, one at the fiducial parameter value, $\vartheta_0$ and the other two at $\pm \Delta \vartheta$, with $\vartheta$ corresponding to each of the parameters $\{T_0, \gamma, \sigma_8, n_{\rm s}\}$ to compute the variation of the mean summary statistic with respect to the model parameters essential for calculating the Fisher information (Eq.~\ref{Eq:Fisher}) and for model parameter estimation (Eq.~\ref{Eq:Estimate}).
 We do this separately for each of the redshifts ($z=2,3$ and 4 for $T_0$ and $\gamma$). In the case of $\sigma_8$ and $n_{\rm s} $, the effect of varying the parameters on Fourier space grows weaker with decreasing redshift. We find that the training doesn't converge with the amount of training data at hand for $z=2$ and that it requires a larger simulated volume to train properly. So for $\sigma_8$ and $n_{\rm s} $, we stick to $z=3,$ and 4 only. The IMNN is then trained over 3,000 sightlines for each parameter value at each redshift value separately. For each of these sightlines, the Gaussian noise added to the transmitted flux is derived from the SNR$_{Fid}$ distribution mentioned earlier. Since the Fourier transformed \lya\ spectra are quite noisy individually, we perform a running mean over 5 pixels in the Fourier profile. We find that this makes the training much more stable without the loss of significant information. We train a separate network for each parameter and each redshift. 
 
 Various training parameters like learning rates, the structure of the network, etc. are mentioned in Table.~\ref{Tab:Training_param} in the appendix. The choice of the training parameters is based on trial and run. We also run the training with different initial random seeds (30 in total) for the neural network to make the exercise more robust and then use all the trained neural estimates on our testing set to estimate the parameters. The parameter estimation is done using the combined results of all the trained neural networks. Examples of a single training instance at $z=3.0$ and the evolution of Fisher information with the number of iterations in training are shown in Fig.~\ref{Fig:Training}. Currently, we have been running the training on CPU and each training instance takes under 1 CPU-hour. We will eventually utilize GPUs for training the IMNN over multiple parameters space, where we expect it to be dramatically faster.

 \begin{figure*}
\centering

    \includegraphics[viewport=20 0 400 200,width=9cm, clip=true]{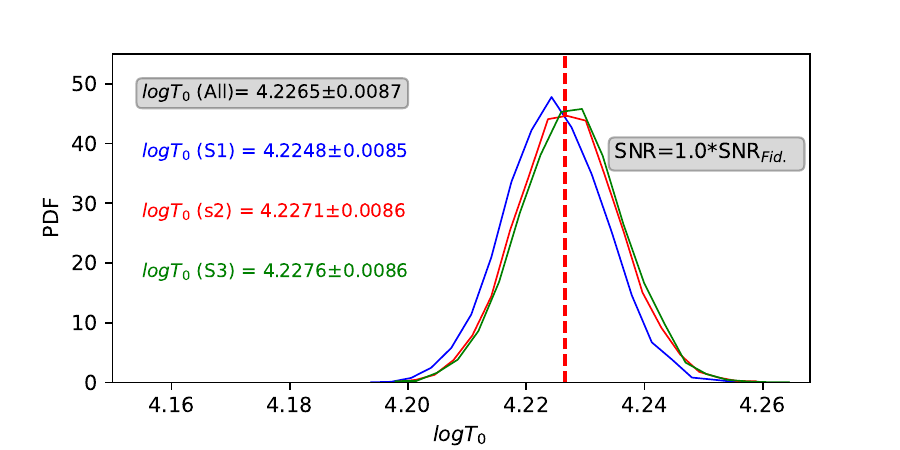}%
    \includegraphics[viewport=20 0 400 200,width=9cm, clip=true]{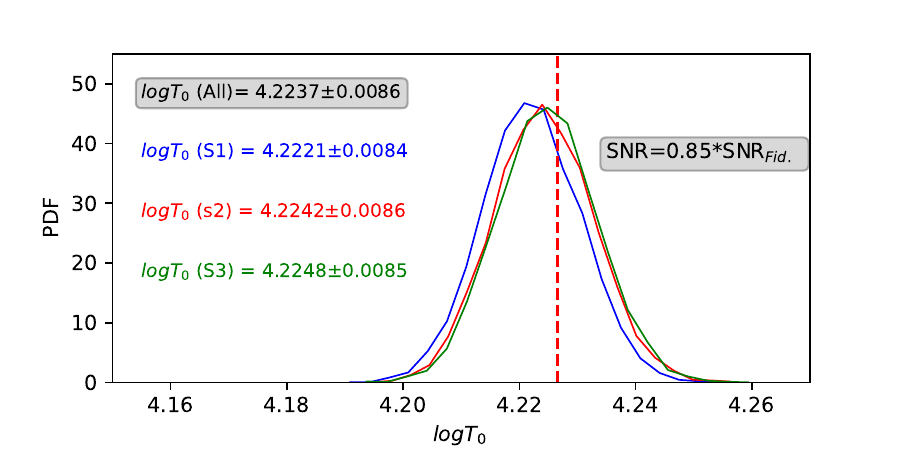}
    
    \includegraphics[viewport=20 0 400 200,width=9cm, clip=true]{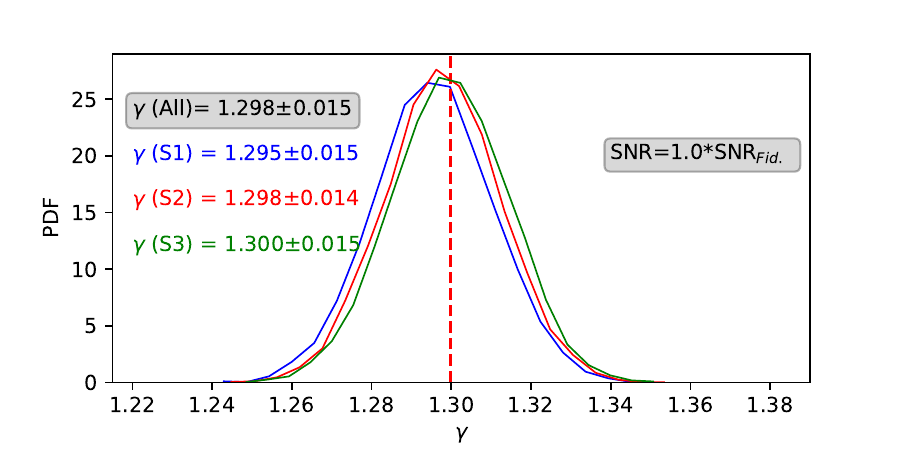}%
    \includegraphics[viewport=20 0 400 200,width=9cm, clip=true]{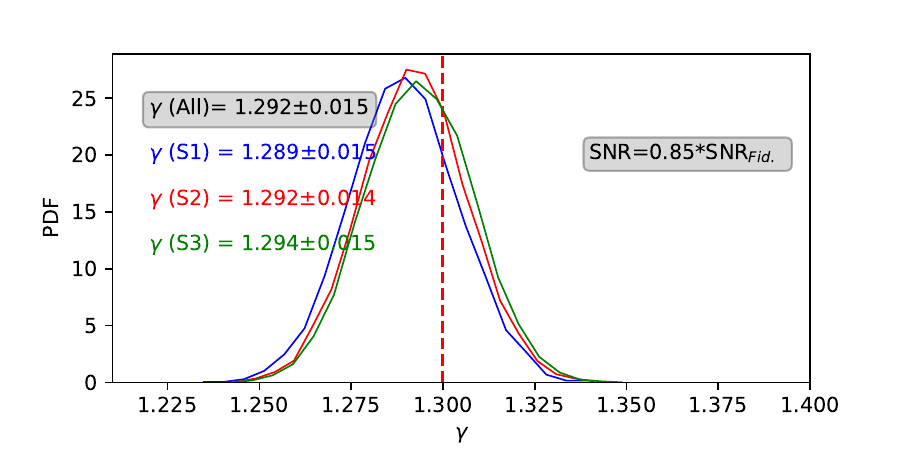}

    \includegraphics[viewport=20 0 400 200,width=9cm, clip=true]{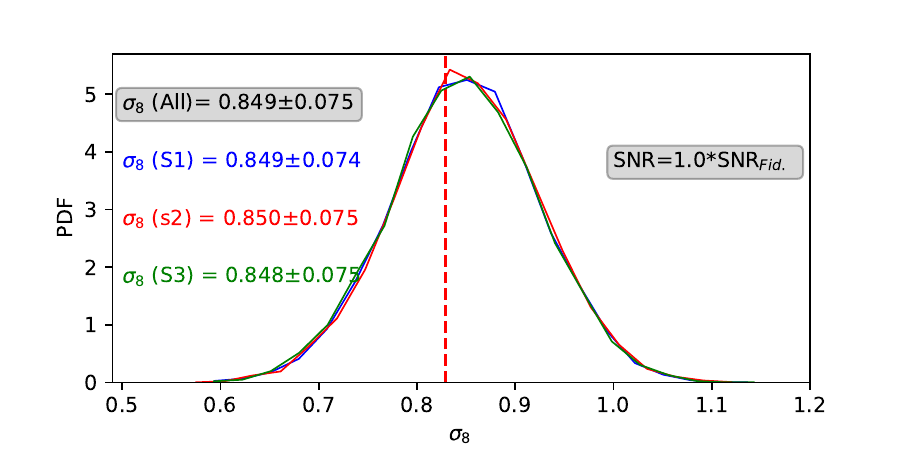}%
    \includegraphics[viewport=20 0 400 200,width=9cm, clip=true]{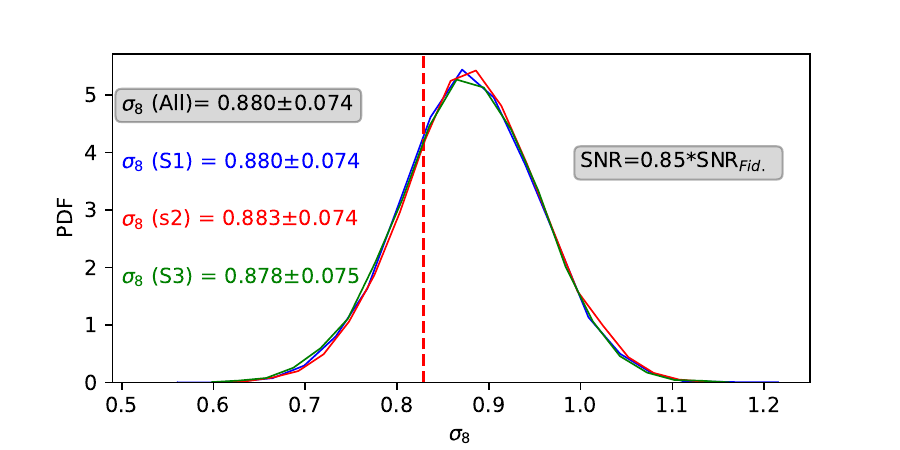}

    \includegraphics[viewport=20 0 400 200,width=9cm, clip=true]{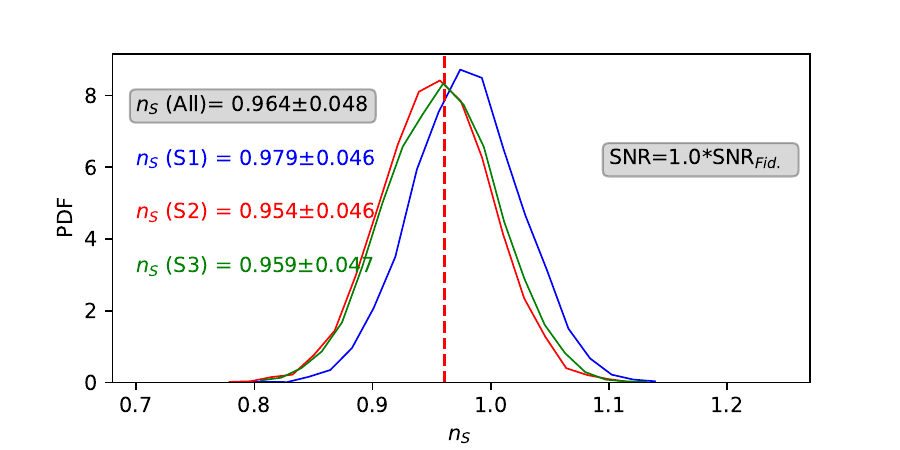}%
    \includegraphics[viewport=20 0 400 200,width=9cm, clip=true]{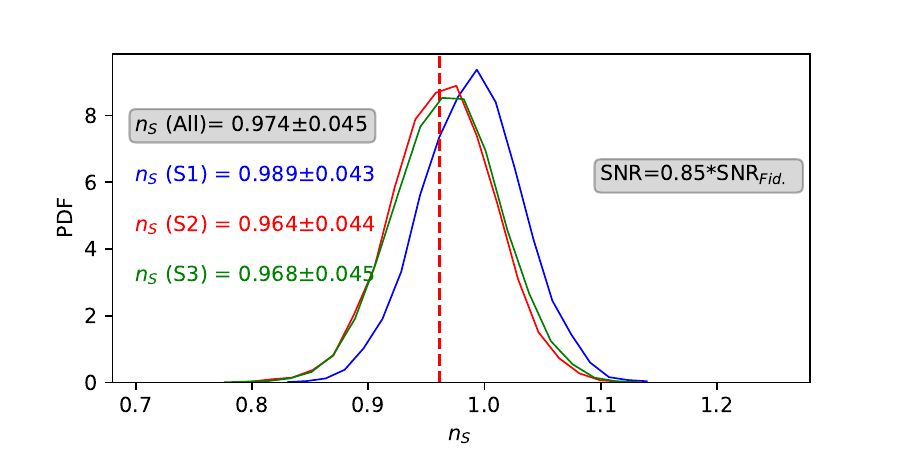}
    
    \caption{Estimation of $T_0, \gamma$, $\sigma_8$ and $n_{\rm s} $ using score-compression from IMNN on the Fourier transformed \lya\ forest transmitted flux for 3 simulations run at fiducial parameter values with different initial random seeds. Each realization in the distribution of the estimated parameters is a bootstrap realization of a sample size of 50 mock spectra over the entire testing set. On the left panels, we plot the estimation for simulated sightlines having SNR distribution SNR$_{Fid}$ with IMNN trained with sightlines having SNR$_{Fid}$. On the right panels, we plot the estimation for $0.85\times$SNR$_{Fid}$ with IMNN trained with SNR$_{Fid}$ sightlines. This shows the sensitivity of the trained neural networks to noise levels deviations in the data.} 
\label{Fig:Param_estimationz=3}
\end{figure*}

\begin{figure}
\centering

    \includegraphics[viewport=20 0 400 200,width=8.5cm, clip=true]{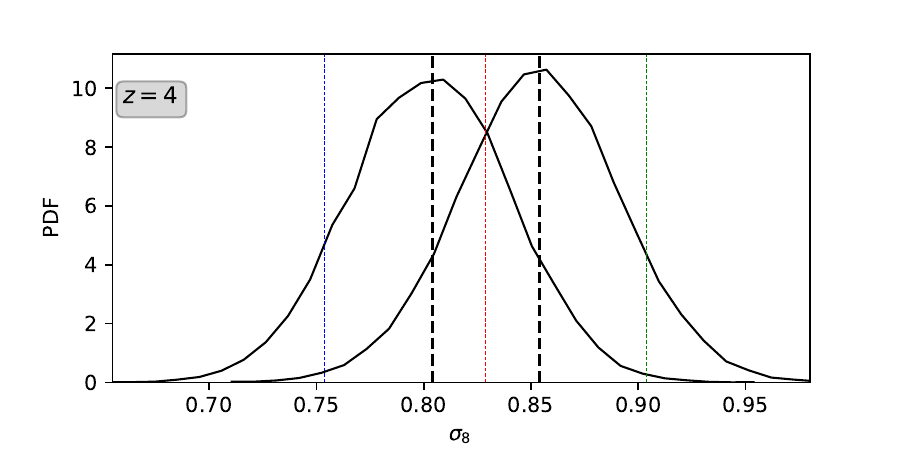}
    
    \includegraphics[viewport=20 0 400 200,width=8.5cm, clip=true]{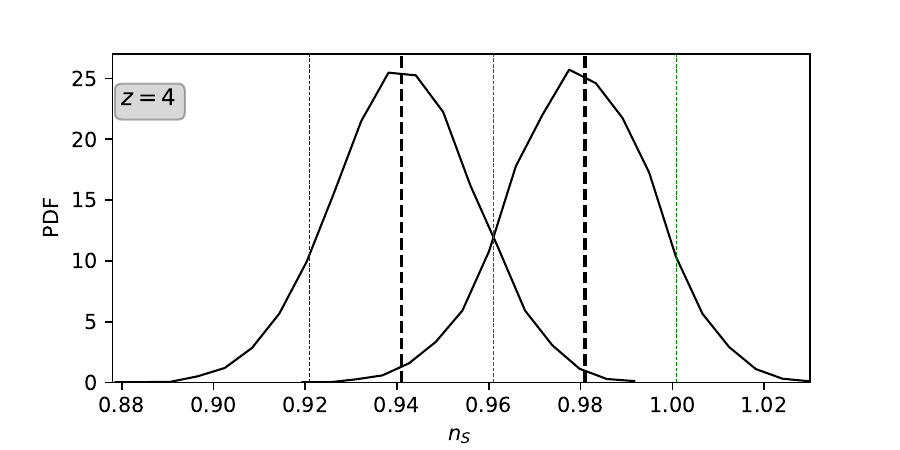}

    \caption{Estimation of $\sigma_8$ and $n_{\rm s} $ at $z=4$ for parameter values (2 black dashed lines) different than the parameter values with which the IMNN has been trained (blue, red, and green dashed lines). The black curves show the PDF of the estimated parameter values (at 2 different values). Each realization in the distribution of the estimated parameters is a bootstrap realization of a sample size of 50 mock spectra over the entire testing set. } 
\label{Fig:Param_estimation_interpolate}
\end{figure}

\section{Parameter estimation with IMNN}\label{Sec:Param_estimate}

\begin{sidewaystable*}
\centering
\caption{Estimation of parameters (for different redshift ranges).}
\begin{tabular}{lllllllll}
\hline
\textbf{Parameter} & \textbf{$z$} & \textbf{Input}  & \textbf{MCMC estimates}  & \textbf{MCMC estimates }  & \textbf{IMNN estimates} & \textbf{IMNN estimates} & \textbf{SNR deviation} & \textbf{Estimate error}  \\
 & & \textbf{Parameter} & (SNR$_{Fid}$) & (0.85$\times$SNR$_{Fid}$) & (SNR$_{Fid}$) & (0.85$\times$SNR$_{Fid}$) & \textbf{Ratios}  & \textbf{Ratios} \\
 & & & & & & & (MCMC/IMNN) & (MCMC/IMNN) \\
\hline
\\
 $\log(T_0)$& $2.0$ & 4.040 & $4.0355 \pm 0.0246$ &  $4.0189\pm 0.0254$ & $4.0394\pm 0.0130$ & $4.0367 \pm 0.0125$ & 6.25 & 1.89 \\
 ($4.040 \pm 0.176$)&  & &  & & & & &  \\
 \\
 $\gamma$& $2.0$ & 1.3 & $1.294 \pm 0.027$ &  $1.274\pm 0.028$ & $1.299\pm 0.018$ & $1.297 \pm 0.0185$ & 10.0 & 1.50 \\
 ($1.3 \pm 0.3$)& & &  &  & & & &\\
 \hline
 \\
  $\log(T_0)$& $3.0$ & 4.225 & $4.223 \pm 0.013$ &  $4.2109\pm 0.0134$ & $4.2265 \pm 0.0087$ & $4.2237 \pm 0.0086$ & 4.17 & 1.52 \\
 ($4.225 \pm 0.176$)&  & &  & & & & &  \\
 \\
 $\gamma$& $3.0$ & 1.3 & $1.295 \pm 0.019$ &  $1.278\pm 0.019$ & $1.298\pm 0.015$ & $1.292 \pm 0.015$ & 2.86 & 1.27 \\
 ($1.3 \pm 0.3$)& & &  &  & & \\
 \\
 $\sigma_8$& $3.0$ & 0.83 & $0.850\pm 0.079$ &  $0.892\pm 0.076$ & $0.849\pm 0.075$ & $0.880 \pm 0.074$ & 1.41 &  1.05   \\
 ($0.83\pm 0.075$)& & &  &  & & & &  \\
 \\
 $n_{\rm s}$& $3.0$ & 0.961 & $0.968\pm 0.050$ &  $0.990\pm 0.049$ & $0.964\pm 0.048$ & $0.974 \pm 0.045$ & 2.2 & 1.04    \\
 ($0.961\pm 0.040$)& & &  &  & & & &  \\
 \hline
 \\
 $\log(T_0)$& $4.0$ & 4.060 & $4.0615 \pm 0.0138$ &  $4.0569\pm 0.0137$ & $4.0631\pm 0.0114$ & $4.0584 \pm 0.0114$ & 0.98 & 1.21 \\
 ($4.060 \pm 0.176$)&  & &  & & & & &  \\
 \\
 $\gamma$& $3.0$ & 1.3 & $1.306\pm 0.029$ & $1.293\pm 0.030$  & $1.298\pm 0.022$ & $1.285\pm 0.022$ & 0.93 &1.32 \\
 ($1.3 \pm 0.3$)& & &  &  & & \\
 \\
 $\sigma_8$& $4.0$ & 0.83 & $0.825\pm 0.048$  & $0.843\pm 0.049$  & $0.830\pm 0.043$ &  $0.848\pm 0.045$ & 1.0 & 1.12 \\
 ($0.83\pm 0.075$)& & &  &  & &  \\
 \\
 $n_{\rm s}$& $3.0$ & 0.961 & $0.957\pm 0.018$ &  $0.96\pm 0.019$ & $0.955\pm 0.016$ & $0.957 \pm 0.016$ & 1.5 &   1.13  \\
 ($0.961\pm 0.040$)& & &  &  & & & &  \\
 \hline
\end{tabular}
\label{Tab:Estimation_z}
\end{sidewaystable*}

As a testing set for the parameter estimation, we use 3 different simulations run at the same fiducial parameter values used for training, but with different initial random seeds, therefore showing the network sightlines which it never saw in training. Currently, we do the testing at the fiducial values of parameter since we only have access to such simulations that have different initial seeds than the ones we trained the neural network on. In the left panels of Fig.~\ref{Fig:Param_estimationz=3}, we show the distribution of the estimated parameters at $z=3.0$ computed using (Eq.~\ref{Eq:Estimate}) for the testing set of sightlines with SNR values derived from SNR$_{Fid}$ distribution. Each realization in the distribution of the estimated parameters is a bootstrap realization of a sample size of 50 mock spectra over the entire validation set. The mean and the associated standard deviation of the estimated parameters are shown in Table.~\ref{Tab:Estimation_z} for all the 3 simulations. The mean and associated error estimates over all the 3 simulations (taking into account the effect of cosmic variance on the error estimates) are also quoted in the plots along with Table.~\ref{Tab:Estimation_z}. Comparing these estimates with those from the MCMC approach using the power spectrum, we find that the error estimates with IMNN are 1.52 times smaller in the case of $\log T_0$ and $1.27$ times in the case of $\gamma$. However, we do not see any significant improvements in the $\sigma_8$ and $n_{\rm s} $ estimates with IMNN. The reason is that varying $\sigma_8$ primarily changes the total power in Fourier space and thus the overall amplitude of the power spectrum. In this case, the MCMC approach with power spectrum only performs well enough to extract nearly maximal information from the Fourier space.

However, it is worth noting that even in the cases where IMNN performs on par with the standard MCMC approach, it offers a significant boost in speed. Once trained, summarizing a new dataset and estimating the model parameters from it is almost an instantaneous process in comparison to MCMC. Thus, using neural network-based approaches provides a substantial advantage in dealing with large astrophysical or cosmological datasets.

\subsection{Effect of noise levels}
Next, we check the robustness of our method against variations in noise levels. Similar to what we did for the MCMC approach with power spectrum, we use the neural network trained on sightlines with noise levels derived from SNR$_{Fid}$ distribution and then test it on sightlines with $0.85\times$SNR$_{Fid}$. The corresponding estimated parameter values are shown in the right panels of Fig.~\ref{Fig:Param_estimationz=3} and also quoted in Table.~\ref{Tab:Estimation_z}. Comparing the deviations with those from the MCMC approach , we find that IMNN estimates are 6.25 and 10.0 times (ratio of the amount of deviation of parameter estimates at $0.85\times$SNR$_{Fid}$ in comparison to SNR$_{Fid}$) less sensitive to variations in noise levels for $\log T_0$ and $\gamma$ at $z=2$, respectively. At $z=3$, the IMNN estimates are less sensitive to noise variations by a factor of 4.17, 2.86, 1.41, and 2.2 times in the case of $\log T_0$, $\gamma$, $\sigma_8$ and $n_{\rm s} $, respectively. At $z=4$, the dependence on noise variations is almost similar between the MCMC and IMNN approaches. In the case of $n_{\rm s}$, though it might seem like there is an improvement (with SNR deviation ratio of 1.5), it might be worth noting that there is not much variation seen in $n_{\rm s} $ estimate itself with noise. In short, we can conclude that the IMNN estimates are more robust against noise variations in comparison to the MCMC approach at $z=2$ and 3, with the approach being more robust at lower redshifts.

\subsection{Effects of instrumental smoothing}

 \begin{table*}
 \centering

\caption{Estimation of parameters for different instrumental smoothing in both training and testing set (gaussian; $z=3.0$).}
\begin{tabular}{llllll}
\hline
\textbf{Parameter}  & \textbf{Input Parameter}  &\multicolumn{4}{c}{\textbf{Output parameter (Error corresponding to 30 sightlines)}}   \\
 &   & No smoothing & FWHM=6\kms\ & FWHM=50\kms\ & FWHM=150\kms\   \\
\hline
\\
 $\log(T_0)$ (IMNN)  & 4.225 & $4.2265\pm 0.0087$ & $4.2237\pm 0.0088$  & $4.2230\pm 0.0360$  & $4.3322\pm 0.4573$ \\
 ($4.225 \pm 0.176$)&   &  &  & &   \\
 \\
 $\log(T_0)$ (MCMC)  & 4.225 & $4.2228\pm 0.0131$ & $4.2226\pm 0.0135$  & $4.2277\pm 0.0325$  & $4.698\pm 0.2260$ \\
 ($4.225 \pm 0.176$)&   &  &  & &   \\
 \hline
 \\
 $\gamma$ (IMNN)   & 1.3 &  $1.298\pm 0.015$ & $1.297\pm 0.015$ & $1.297\pm 0.040$ & $1.268\pm 0.168$  \\
 ($1.3 \pm 0.3$)& &   &  & &   \\
 \\
 $\gamma$ (MCMC)   & 1.3 &  $1.295\pm 0.019$ & $1.297\pm 0.020$ & $1.301\pm 0.041$ & $1.279\pm 0.162$  \\
 ($1.3 \pm 0.3$)& &   &  & &   \\
 \hline
 \\
 $\sigma_8$ (IMNN)  & 0.83 & $0.849\pm 0.075$  & $0.850\pm 0.075$  & $0.835\pm 0.084$ & $0.861\pm 0.085$ \\
 ($0.83\pm 0.075$)& &   &  & &     \\
  \\
 $\sigma_8$ (MCMC)  & 0.83 & $0.850\pm 0.079$  & $0.844\pm 0.080$  & $0.848\pm 0.085$ & $0.848\pm 0.099$ \\
 ($0.83\pm 0.075$)& &   &  & &     \\
 \hline
 \\
 $n_{\rm s}$ (IMNN)  & 0.961 & $0.964\pm 0.048$  & $0.966\pm 0.047$  & $0.966\pm 0.049$ & $0.972 \pm 0.70$ \\
 ($0.961\pm 0.040$)& &   &  & &     \\
  \\
 $n_{\rm s}$ (MCMC)  & 0.961 & $0.968\pm 0.050$  & $0.969\pm 0.050$  & $0.963\pm 0.058$ & $0.966\pm 0.064$ \\
 ($0.961\pm 0.040$)& &   &  & &     \\
 \hline
 \end{tabular}
 \label{Tab:Estimation_smoothing}
 \end{table*}
 
 We investigate the effects of instrumental smoothing in the parameter estimation procedure by training the neural networks with sightlines convolved with a Gaussian profile. We use Gaussian profiles having FWHM=6, 50, and 150 \kms\ to simulate the usual convolutional scales of high, medium and low resolution \lya\ forest spectra. In Table.~\ref{Tab:Estimation_smoothing}, we present the results of the effects of instrumental smoothing on the estimation of parameters with the IMNN approach and compare it with the results from the MCMC approach. For the parameters $T_0$ and $\gamma$, we find that the estimation with IMNN is better with high-resolution spectra having FWHM=6\kms. The estimates with FWHM=6\kms are roughly similar to what we obtained with unsmoothed sightlines. In the case of FWHM=50 and 150\kms, we do not see any improvements in the parameter estimates using IMNN over MCMC with power spectrum. In fact, the parameter estimation fails for $T_0$ and $\gamma$ at FWHM=150\kms, as can seen by the large errorbars. This is expected as the thermal effects on \lya\ forest are small-scale effects that alter the broadening of \lya\ absorption lines. Large-scale smoothing introduced by low-resolution spectrographs washes away this small-scale information. In conclusion, the improvements that we observed with IMNN over MCMC with power spectrum come from enhanced extraction of small-scale information by the IMNN, which can only be realized with high-resolution spectra. 

 On the other hand, $\sigma_8$ and $n_{\rm s} $ primarily affect the global amplitude of the power spectrum, as mentioned earlier. So, we do not find any appreciable change in the parameter estimates by introducing Gaussian convolution with FWHM=6, 50, and 150 \kms. The IMNN approach works on par with MCMC using power spectrum for all the cases.

 \subsection{Effect of continuum uncertainty}

 \begin{table*}
\centering
\caption{Effects of continuum variation on the estimation of astrophysical parameters $log T_0$ and $\gamma$ at $z=3.0$. In the case of cosmological parameters $\sigma_8$ and $n_S$, IMNN fails to extract continuum-independent information out of the Fourier space. }
\begin{tabular}{llll}
\hline
\textbf{Parameter}  & \textbf{Input Parameter}  &\multicolumn{2}{c}{\textbf{Output parameter (Error corresponding to 30 sightlines)}}   \\
 &   & Continuum ($\mu=1.0$, $\sigma=0$) & Continuum ($\mu=0.9$, $\sigma=0.1$)  \\
\hline
\\
 $\log(T_0)$ (IMNN)  & 4.225 & $4.224\pm 0.009$ & $4.224\pm 0.010$  \\
 ($4.225 \pm 0.176$)&   &  &  \\ 
 \\
 $\log(T_0)$ (MCMC)  & 4.225 & $4.223\pm 0.013$ & $4.204\pm 0.014$  \\
 ($4.225 \pm 0.176$)&   &  &     \\
 \hline
 \\
 $\gamma$ (IMNN)   & 1.3 &  $1.300\pm 0.015$ & $1.301\pm 0.015$  \\
 ($1.3 \pm 0.3$)& &   &    \\
 \\
 $\gamma$ (MCMC)   & 1.3 &  $1.295\pm 0.019$ & $1.265\pm 0.021$   \\
 ($1.3 \pm 0.3$)& &   &     \\
 \hline
 \end{tabular}
 \label{Tab:Estimation_continuum}
 \end{table*}

In this section, we check the effectiveness of IMNN in extracting continuum-independent information from the Fourier space for parameter estimation. For this, we first generate a training set of mock \lya\ sightlines with their continuum levels altered. We generate random continuum levels for each sightline from a Gaussian distribution of mean $\mu=1$ and standard deviation$\sigma=0.2$ and then normalize the transmitted flux for each sightline to these random continuum levels. The validation set is also treated in the same way. The neural network is then trained on these continuum-altered \lya\ transmitted flux sightlines. We then estimate the parameters using these trained neural networks for sightlines having continuum levels of ($\mu=1$, $\sigma=0$) and ($\mu=0.9$, $\sigma=0.1$). The estimated astrophysical parameters $log T_0$ and $\gamma$ are presented in Table.~\ref{Tab:Estimation_continuum} at $z=3$. We also compare our results with the standard MCMC approach for parameter estimation using the power spectrum. We find that training the IMNN on sightlines having varying continuum levels makes it able to extract continuum-independent information. The estimated $log T_0$ values for ($\mu=1$, $\sigma=0$) and ($\mu=0.9$, $\sigma=0.1$) sightlines using IMNN are $4.224\pm 0.009$ and $4.224\pm 0.010$, respectively, as opposed to $4.223\pm 0.013$ and $4.204\pm 0.014$ using standard MCMC approach. In the case of $\gamma$, the estimated parameters using IMNN are $1.300\pm 0.015$ and $1.301\pm 0.015$ as opposed to $1.295\pm 0.019$ and $1.265\pm 0.021$ using MCMC approach. So, IMNN clearly outperforms the MCMC approach when considering robustness against continuum uncertainties in estimating $T_0$ and $\gamma$.

The situation is different in the case of the cosmological parameters $\sigma_8$ and $n_S$ which are known to be sensitive to the global amplitude of the power spectrum, as opposed to the astrophysical parameters $T_0$ and $\gamma$ which are more sensitive to the small scale information. As expected, the IMNN trained on ($\mu=1.0$, $\sigma=0.2$) sightlines doesn't estimate the parameters accurately for sightlines having ($\mu=1.0$, $\sigma=0.0$) and ($\mu=0.9$, $\sigma=0.1$). In the case of $\sigma_8$, the estimated parameters for these 2 cases are $0.891\pm 0.090$ and $0.887\pm 0.097$, respectively. In the case of $n_S$, the estimated parameters are $0.995\pm 0.050$ and $0.991\pm 0.062$.

\section{Robustness check with different simulation}

To check the robustness of parameter estimation using IMNN on \lya\ forest spectra in the Fourier domain, we use alternate simulations to estimate parameters based on the neural trained with the Sherwood-Relics simulations. For this, we use mock\lya\ forest spectra from the CAMELS project \citep{navarro2021}, generated using (25$h^{-1}$cMpc)$^3$ IllustrisTNG simulation box at $z=2$. We do not perform this exercise at $z=3$ since IGM temperatures are elevated in the case of Sherwood-Relics simulations due to \HeII\ reionization, thus making the trained neural networks unfit for testing on IllustrisTNG simulations that do not incorporate this. The cosmological parameters used in this simulation are (\{$\Omega_m$, $\Omega_b$, $\Omega_\Lambda$, $\sigma_8$, $n_{\rm s} $, $h$\} = \{0.30, 0.049, 0.70, 0.84, 0.9624, 0.6711\}). We first linearly interpolate the spectra to the grid size of our simulation and then add random Gaussian noise with SNR levels drawn from the SNR$_{Fid}$ distribution mentioned before. We then match the mean flux of the sightlines to the mean flux of our fiducial simulation. We simultaneously check the parameter estimation for $\log T_0$ using the standard MCMC approach from the power spectrum with the IMNN approach.

The expected values of $\log T_0$ and $\gamma$ in the CAMELS simulations are approximately 3.97 and 1.26, respectively. In the case of MCMC with power spectrum, the $\log T_0$ predicted is $3.997\pm 0.036$. With IMNN, the value predicted is $3.9898\pm 0.0154$. We see that the estimated values are consistent with each other as well as the expected values from CAMELS within $1\sigma$ errorbars with the IMNN predicting it with $\approx 2.34$ times better accuracy in case of $\log T_0$ and $\approx 1.73$ times better accuracy in case of $\gamma$. We thus conclude that this neural network approach is relatively robust and doesn't learn simulation-specific features from the Fourier transformed \lya\ forest spectra.

\begin{figure*}
\centering

    \includegraphics[width=6.5cm]{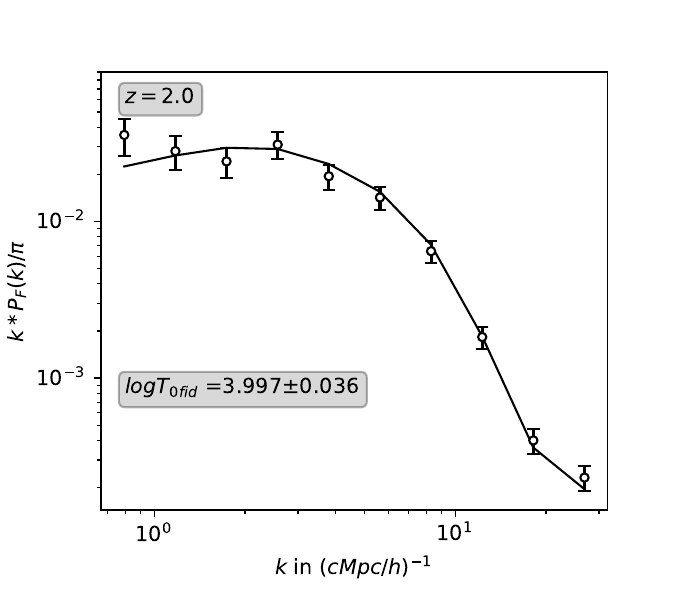}%
    \includegraphics[width=11.5cm]{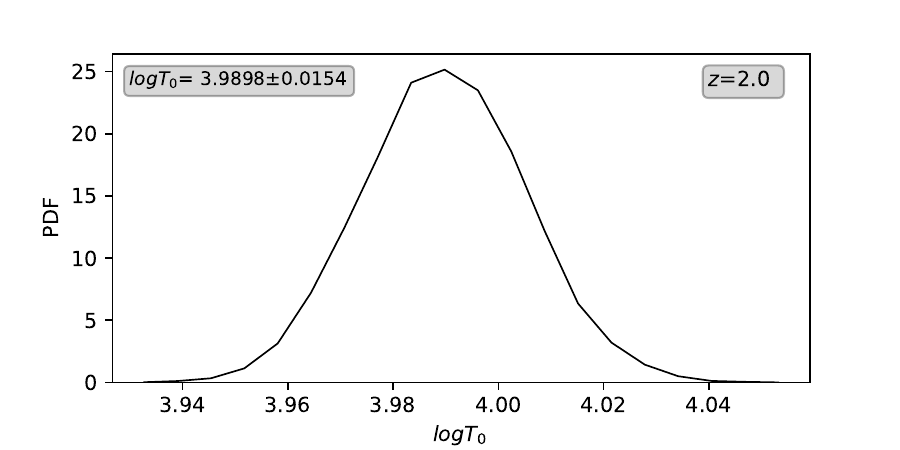}

    \includegraphics[width=6.5cm]{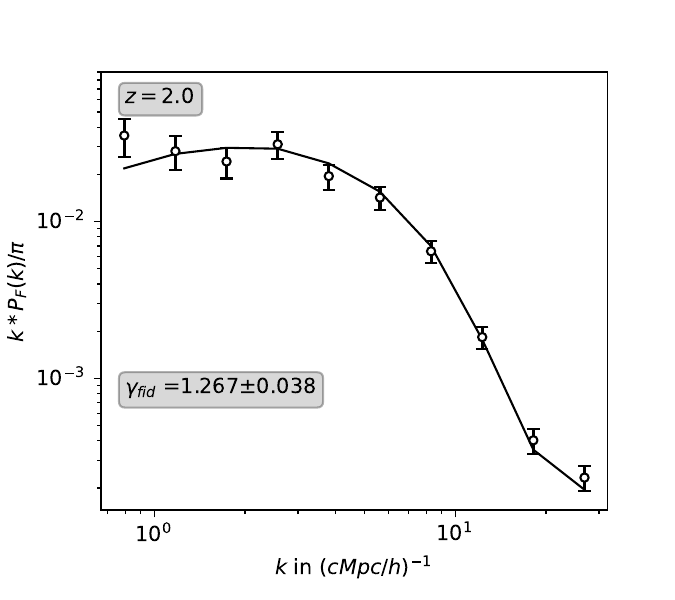}%
    \includegraphics[width=11.5cm]{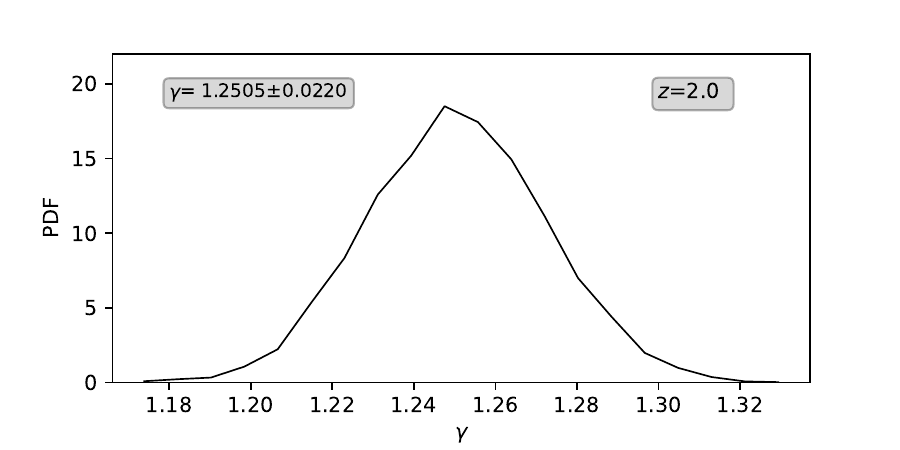}

    \caption{Estimation of $T_0$ and $\gamma$ using MCMC with power spectrum and IMNN on the Fourier transformed \lya\ forest transmitted flux for CAMELS simulation at $z=2$. The expected parameter values are $\log T_0=3.97$ and $\gamma=1.26$. The power spectrum modeling for the MCMC approach and the training for the IMNN approach were done with previous Sherwood-Relics simulations. We check the robustness of both approaches by testing them on a different simulation.} 
\label{Fig:Param_estimationz=2}
\end{figure*}

\section{Parameter Inference in 2d}\label{Sec:Param_2d}

\begin{figure}
\centering

    \includegraphics[viewport=35 0 305 200,width=8cm, clip=true]{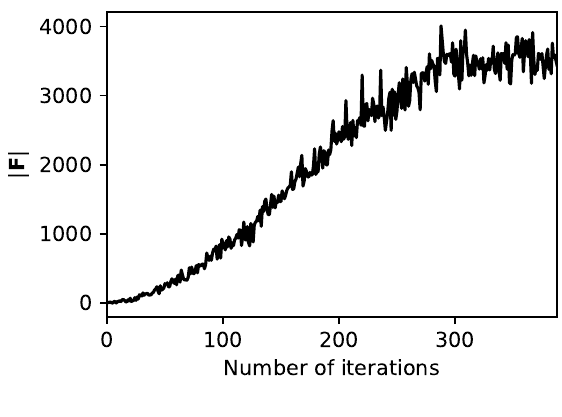}
    
    \includegraphics[width=9cm]{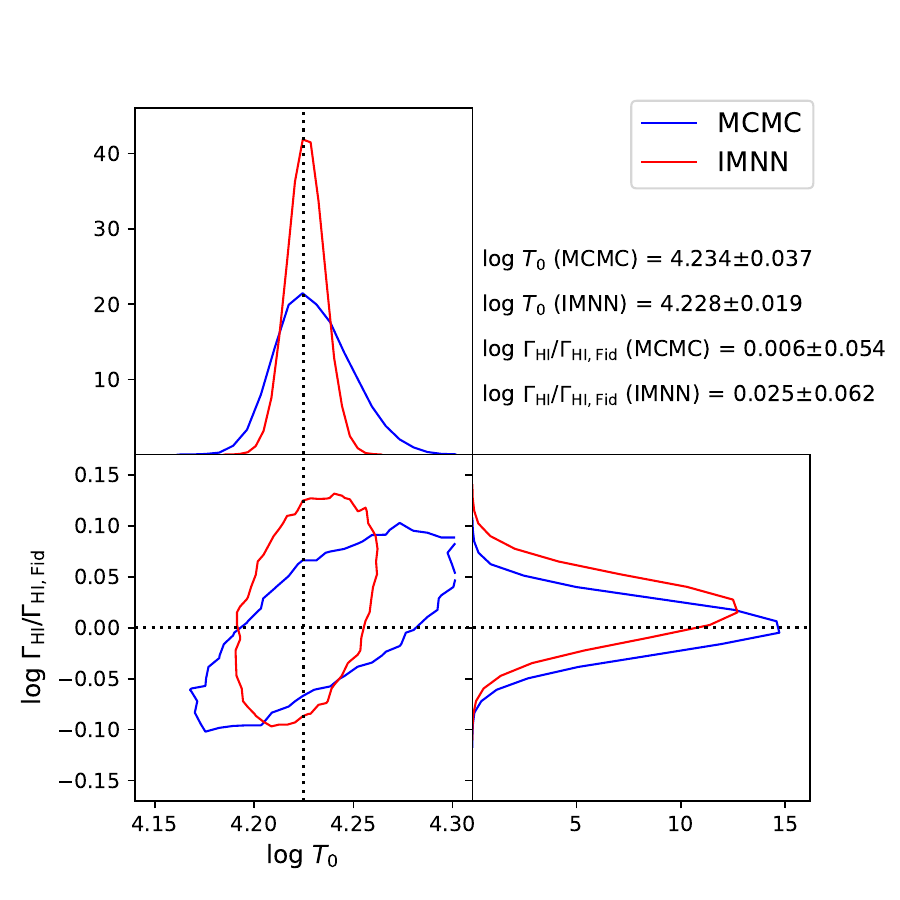}

    \caption{Top: IMNN training plot (using SimulatorIMNN subclass of IMNN) corresponding to thermal parameter $T_0$ and \HI\ photo-ionization rate $\Gamma_{\rm HI}$ for $z=3$. The plot shows the evolution of Fisher information during training with the training epoch. Bottom: Contour plots showing the 68\% credible region for  log$T_0$ and log$\Gamma_{\rm HI}/\Gamma_{\rm HI,Fid}$ estimates based on the IMNN approach and with MCMC. It is to be noted that in the case of MCMC, the contour plot corresponds to a posterior distribution drawn from running MCMC using a flat prior (see Eq.~\ref{Eq:Prior}) while in the case of IMNN, it corresponds to the distribution of the estimates. Accompanying this are the plots for the distribution functions (for MCMC and IMNN, respectively) for both the parameters log$T_0$ and log$\Gamma_{\rm HI}/\Gamma_{\rm HI, Fid}$ individually by marginalizing over the other parameters.} 
\label{Fig:Param_estimation_2d}
\end{figure}

In the previous sections, we used the IMNN approach to estimate thermal and cosmological parameters individually. However, in a realistic scenarios, such estimation would be performed jointly over several parameters, and then individual parameter estimates are obtained by marginalizing over the other parameters. In this section, we demonstrate a 2D parameter estimation. Ideally, we would have liked to do this for the thermal parameters $T_0$ and $\gamma$. However, we find that IMNN doesn't properly learn to perform 2D parameter inference and lift the degeneracy between these two degenerate parameters, due to the lack of training samples where both $T_0$ and $\gamma$ are varied simultaneously. Since we currently lack such simulations in the SHERWOOD suite, as a proof of concept we demonstrate 2D parameter inference for the 2 parameters $T_0$ and \HI\ photo-ionization rate $\Gamma_{\HI}$. 2D parameter inferences $T_0$ and $\gamma$ using several parameter variation instances, similar to \citep{nayak2023}, will be addressed in a future work.

For this work, we use our original 3 simulations at $z=3$ with log$T_0=4.225 \pm 0.176$ and simply generate sightlines corresponding to $\Gamma_{\rm HI}= \Gamma_{\rm HI}/1.25, \Gamma_{\rm HI}$ and $1.25\times \Gamma_{\rm HI}$ to generate 9 simulations in total with varied $T_0$ and $\Gamma_{\rm HI}$. We use the SimulatorIMNN\footnote{\url{https://docs.aquila-consortium.org/imnn/main/pages/modules.html\#simulatorimnn}} subclass for 2D parameter inference as we find that it is more suited for simultaneous parameter variations. It is also suited for future exercises where we will train the neural network over simulations with different parameters in several dimensions. Unlike the NumericalGradientIMNN used in the previous sections --which uses the numerical gradient between the training samples corresponding to the varied parameter values--, SimulatorIMNN requires a function to generate the training samples based on input parameter values. We do this by defining a function that linearly interpolates the Fourier transformed field between the 9 simulations with varied $T_0$ and $\Gamma_{\rm HI}$. Such an interpolation scheme becomes possible in our case since we use simulations having the same initial density fields and differ only in the parameters $T_0$ and $\Gamma_{\rm HI}$. This means that we can easily interpolate in the parameter space between sightlines since they essentially trace similar density fields that differ only in $T_0$ and $\Gamma_{\rm HI}$. Currently, to keep the exercise simple, we avoid modeling the noise effects on the Fourier-transformed fields. We use sightlines having a uniform noise distribution with SNR=50. In order to  compare the IMNN approach with the standard MCMC approach using power spectra, we adopt the same interpolation scheme to model the power spectrum as a function of $T_0$ and $\Gamma_{\rm HI}$, which we use subsequently in the likelihood for the MCMC analysis.

For the training of IMNN, we use 5000 sightlines corresponding to each parameter value shot from the 9 simulation boxes. For SimulatorIMNN, the training does not require a validation set. The network architecture and training hyperparameters used are mentioned in Table~\ref{Tab:Training_param}. We test the trained neural network on 5000 sightlines each corresponding to the 3 simulation boxes run with fiducial parameters but having different initial seed density fields. The training is done at $z=3$ and the evolution of the Fisher information with the number of iterations in training is shown in the left panel of Fig.~\ref{Fig:Param_estimation_2d}.
In the right panel of Fig.~\ref{Fig:Param_estimation_2d}, we make a 68\% confidence limit contour plot for the 2D histogram of the log$T_0$ and log$\Gamma_{\rm HI}/\Gamma{\rm HI, Fid}$ estimates based the IMNN approach and the standard MCMC approach.  Similar to the previous exercises, each estimate corresponds to bootstrap realizations over 50 sightlines for both IMNN and MCMC approaches. Along with this, we also plot the distribution functions for both these parameters individually by marginalizing the other parameters. The average estimates along with the associated error (computed as the width of 68\% confidence interval of the distribution function) are quoted along with the plots. We find that both approaches predict the fiducial parameters (log$T_0=4.225$ and log$\Gamma_{\rm HI}/\Gamma{\rm HI, Fid}=0.0$) well within the errorbars. For log$\Gamma_{\rm HI}/\Gamma{\rm HI, Fid}$, the errors are similar between the two approaches, with the MCMC errorbars being about 13\% smaller with respect to the IMNN errorbars. However, this is expected as mentioned in the previous sections that IMNN gives similar estimates as MCMC when the associated parameter has a more global effect (such as $\Gamma_{\rm HI}$ which uniformly alters the photo-ionizing background at all scales). The IMNN estimates get significantly better for log$T_0$ where the effect is on small scales with MCMC errorbars being 95\% larger with respect to the IMNN errorbars. This exercise here demonstrates a case for 2D parameter estimation using IMNN which outperforms the standard MCMC approach in extracting small-scale information from the \lya\ forest in Fourier space. This also sets a pathway for performing a full N-D parameter estimation for \lya\ forest using IMNN.

\section{Summary and Discussion}
With the current advent of large astrophysical and cosmological data, reducing data dimensionality is now of utmost importance. Reducing the entire dataset to a set of summary statistics relevant to the problem at hand with minimum loss of information is a way to achieve that. One popular approach of summarizing large datasets is the Massively Optimised Parameter Estimation and Data (MOPED; \cite{heavens2000}) which reduces the entire dataset to the number of model parameters describing the data. One can also use neural networks to obtain a mapping function that maps the dataset to a set of non-linear summaries with minimal loss of information. 

In this work, we primarily perform 1D parameter estimation for model parameters $T_0$ (IGM temperature at mean cosmic density), $\gamma$ (slope IGM temperature density $\Delta$ relation $T=T_0\Delta^{\gamma-1}$), $\sigma_8$ and $n_S$ using the Information Maximizing Neural Network (IMNN; \cite{charnock2018}). We perform a 1D estimation since we find that training the neural network to do N-D parameter estimation requires the training set to have simulations where the parameters are varied simultaneously. We currently do not have access to such simulations for the model parameters mentioned, but we do perform a 2D parameter estimation for $T_0$ and \HI\ photo-ionization rate $\Gamma_{\rm HI}$ to demonstrate how one can use IMNN to perform N-D parameter estimation in the future. For the 1D parameter estimation, we use IMNN to summarize the \lya\ forest data in Fourier space as a single summary value with minimal loss of information. We obtain these summary values individually. One can then use the summary values to obtain the model parameter estimates. We then compare this approach with a standard technique of Maximum Likelihood Estimation (MCMC) from the Fourier space using power spectrum. We find that the IMNN approach leads to a significant enhancement in the parameter estimation for $T_0$ and $\gamma$. For $\log T_0$, the enhancements are by a factor of 1.89, 1.52 and 1.21 times at $z=2,3$ and 4, respectively. For $\gamma$, the enhancements are by a factor of 1.5, 1.27, and 1.32. Besides enhancing the estimation of these parameters, we also find that the neural networks are more robust against variation in noise levels in the spectra. We confirm this by performing an exercise where we estimate the parameters using a testing set of spectra having an SNR distribution that is 0.85 times lower than the fiducial distribution of SNR. At $z=2$ and 3, the neural networks were less sensitive to this variation in noise levels than the standard MCMC approach by a factor of 6.25 and 4.17 times in the case of $\log T_0$ and by a factor of 10.0 and 2.86 times for $\gamma$. However, at $z=4$, the sensitivity to noise variations was more or less on par with the MCMC approach. We also performed the parameter estimation using IMNN with the cosmological parameters $\sigma_8$ and $n_S$. We find that the estimates and their sensitivity to noise variations were on par with the MCMC estimates. We think that this might be because varying $\sigma_8$ or $n_S$ primarily varies the total power in the Fourier space and thus, the global amplitude of the power spectrum. Hence, the MCMC approach with the power spectrum is good enough to extract maximal information present in the Fourier space. However, it is worth emphasizing here that even in cases where a standard MCMC approach with power spectrum works well enough, a trained neural network can offer a significant boost in speed in summarizing a new dataset and estimating the model parameters from that. This becomes especially important in the current era of large astrophysical and cosmological dataset.

We also find that IMNN can provide parameter estimates for $T_0$ and $\gamma$ which are more robust against continuum uncertainties in comparison to the standard MCMC estimates. It does this by extracting continuum-independent small-scale information from the Fourier domain for these parameters. It, however, fails to do this for the cosmological parameters $\sigma_8$ and $n_S$ which are more sensitive to the global amplitude of the power spectrum.

Additionally, we also checked the parameter estimation process with different instrumental smoothing scales with FWHM=6, 50 and 150 \kms corresponding to typical high, moderate and low-resolution spectra. Interestingly, we find that the improvements seen in estimating $T_0$ and $\gamma$ are also seen for FWHM=6\kms, but not for 50 and 150\kms. In fact, they perform on par with the MCMC estimates. Based on this, we can infer that the improvement we see in the case of IMNN comes mostly from additional small-scale information. In conclusion, IMNN works efficiently in extracting the maximal small-scale information in comparison to the MCMC approach. In doing so, it not only doesn't compromise on robustness against noise level variations but rather improves it.

We also demonstrate a 2D parameter estimation for model parameters $T_0$ and $\Gamma_{\rm HI}$ using IMNN. In doing so, we find that the IMNN estimation works on par with MCMC approach using power spectrum for $\Gamma_{\rm HI}$ (with MCMC estimates having about 13\% smaller errors), but we see significant improvements in $log T_0$ estimates wherein the errors are about 84\% smaller. This behavior is similar to what we have seen before. Varying $\Gamma_{\rm HI}$ uniformly rescales the optical depth field of \lya\ forest and hence induces global effect. For such parameters, MCMC approach with power spectrum can extract maximal information. The IMNN approach outperforms the MCMC approach for $T_0$ (which induces small-scale effects on the \lya\ forest) by extracting additional small-scale information.

In our upcoming work, we will implement this on observational data comprising of high resolution \lya\ forest spectra obtained from publicly available surveys like KODIAQ \citep{omeara2017} based on KECK/HIRES spectrograph and SQUAD survey \citep{murphy2019} based on VLT/UVES. This will also be complemented with high-resolution spectra from ESPRESSO to improve upon the current estimates of the thermal evolution of the IGM.

\section*{Acknowledgements}
SM, SC and GC acknowledge financial support of the Italian Ministry of University
and Research with PRIN 201278X4FL, PRIN INAF 2019 "New Light on the Intergalactic Medium" 
and the ‘Progetti Premiali’ funding scheme. MV, SC, RT are supported by INDARK INFN PD51 grant. RT acknowledges co-funding from Next Generation EU, in the context of the National Recovery and Resilience Plan, Investment PE1 – Project FAIR Future Artificial Intelligence Research''. This resource was co-financed by the Next Generation EU [DM 1555 del 11.10.22]. RT is partially supported by the Fondazione ICSC, Spoke 3 ``Astrophysics and Cosmos Observations'', Piano Nazionale di Ripresa e Resilienza Project ID CN00000013 ``Italian Research Center on High-Performance Computing, Big Data and Quantum Computing'' funded by MUR Missione 4 Componente 2 Investimento 1.4: Potenziamento strutture di ricerca e creazione di ``campioni nazionali di R\&S (M4C2-19 )'' - Next Generation EU (NGEU). SM would also like to acknowledge Valentina D'Odorico and Prakash Gaikwad for valuable discussions, insights, and feedback regarding the topic at hand.



\bibliographystyle{aa}
\bibliography{main} % if your bibtex file is called example.bib

%\begin{thebibliography}{99}
%\end{thebibliography}

%%%%%%%%%%%%%%%%%%%%%%%%%%%%%%%%%%%%%%%%%%%%%%%%%%

%%%%%%%%%%%%%%%%% APPENDICES %%%%%%%%%%%%%%%%%%%%%

\appendix

\section{Network architecture}

\begin{table*}
\centering
\caption{Training parameters used for each network.}
\begin{tabular}{llllll}
\hline
\textbf{Parameter} & \textbf{Redshift} & \textbf{No. summaries}  & \textbf{Network description}  & \textbf{Learning rate} & $\lambda,\ \epsilon$   \\
\hline
\\
$\log(T_0)$ [1D] & 2.0 & 1 & $3\times 256$ &  0.005 & 10.0, 0.1  \\
 $\gamma$ [1D]& 2.0 & 1 & $2\times 128$ &  0.01 & 10.0, 0.1  \\
 \hline
\\
$\log(T_0)$ [1D]& 3.0 & 1 & $2\times 256$ &  0.01 & 10.0, 0.1  \\
 $\gamma$ [1D]& 3.0 & 1 & $2\times 128$ &  0.01 & 10.0, 0.1  \\
 $\sigma_8 [1D]$ & 3.0 & 1 & $2\times 512$ &  0.005 & 10.0, 0.1  \\ 
 $n_S$ [1D] & 3.0 & 1 & $4\times 256$ &  0.0001 & 10.0, 0.1  \\ 
 $\log(T_0), \Gamma_{\rm HI}$ [2D]& 3.0 & 2 & $2\times 512$ &  0.001 & 10.0, 0.1  \\
 \hline
 \\
$\log(T_0)$ [1D]& 4.0 & 1 & $2\times 256$ &  0.005 & 10.0, 0.1  \\
$\gamma$ [1D]& 4.0 & 1 & $3\times 128$ &  0.001 & 10.0, 0.1  \\
 $\sigma_8$ [1D] & 4.0 & 1 & $2\times 256$ &  0.01 & 10.0, 0.1  \\ 
 $n_S$ [1D] & 4.0 & 1 & $3\times 256$ &  0.001 & 10.0, 0.1  \\ 
 \hline
\end{tabular}
\label{Tab:Training_param}
\end{table*}

The details of the network parameters including the number of layers and nodes in each layer, the learning rate and regularization parameters ($\lambda$ and $\epsilon$) are given in Table.~\ref{Tab:Training_param}. We have used an additional layer in case of $\log T_0$ using which we input the log of the $\sqrt{k}\delta_F(k)$. We find that this gives a better estimation of $T_0$. Also, we consider the summary covariances over a batch of 2000 sightlines in case of $\log T_0$ and $\gamma$ and 3000 in case of $\sigma_8$. As a stopping criterion of the training, we choose a patience value of 20, which is the number of iterations where there is no increase in the value of the determinant of the Fisher information matrix.

%%%%%%%%%%%%%%%%%%%%%%%%%%%%%%%%%%%%%%%%%%%%%%%%%%

% Don't change these lines
\end{document}